\documentclass[linenumber]{aastex631}

\shortauthors{Deng et al.}
\graphicspath{{./}{figures/}}
\usepackage{amsmath}
\usepackage{longtable}
\usepackage{graphicx}
\usepackage{xcolor}
\makeatletter

\newcommand{\Rmnum}[1]{\expandafter\@slowromancap\romannumeral #1@}
\makeatother
\begin{document}
\title{Modeling the Multi-Wavelength Afterglow of Short Gamma-Ray Bursts with a Plateau Phase}
\correspondingauthor{Yong-Feng Huang, Chen Deng}
\email{hyf@nju.edu.cn, dengchen@nju.edu.cn}

\author[0000-0002-2191-7286]{Chen Deng}
\affiliation{School of Astronomy and Space Science, Nanjing
University, Nanjing 210023, China}
\affiliation{Key Laboratory of Modern Astronomy and Astrophysics
(Nanjing University), Ministry of Education, Nanjing 210023, China}

\author[0000-0001-7199-2906]{Yong-Feng Huang}
\affiliation{School of Astronomy and Space Science, Nanjing
University, Nanjing 210023, China}
\affiliation{Key Laboratory of Modern Astronomy and Astrophysics
(Nanjing University), Ministry of Education, Nanjing 210023, China}

\author[0000-0002-2162-0378]{Abdusattar Kurban}
\affiliation{Xinjiang Astronomical Observatory, Chinese Academy of Sciences, Urumqi 830011, Xinjiang, China}
\affiliation{Xinjiang Key Laboratory of Radio Astrophysics, Urumqi 830011, Xinjiang, China}
\affiliation{State Key Laboratory of Radio Astronomy and Technology, A20 Datun Road, Chaoyang District, Beijing, 100101, China}

\author[0000-0001-9648-7295]{Jin-Jun Geng}
\affiliation{Purple Mountain Observatory, Chinese Academy of Sciences,
Nanjing 210023, China}

\author[0000-0001-7943-4685]{Fan Xu}
\affiliation{Institute of Space Weather, Nanjing University of Information 
Science and Technology, Nanjing, China}

\author[0009-0000-0467-0050]{Xiao-Fei Dong}
\affiliation{School of Astronomy and Space Science, Nanjing
University, Nanjing 210023, China}

\author[0000-0001-7892-9790]{Hao-Xuan Gao}
\affiliation{Purple Mountain Observatory, Chinese Academy of Sciences,
Nanjing 210023, China}

\author[0000-0002-7044-733X]{En-Wei Liang}
\affiliation{Guangxi Key Laboratory for Relativistic Astrophysics, School
of Physical Science and Technology, \\Guangxi University, Nanning 530004, China}

\author[0000-0002-1343-3089]{Liang Li}
\affiliation{Institute of Fundamental Physics and Quantum Technology,
Ningbo University, Ningbo, Zhejiang 315211, China}
\affiliation{School of Physical Science and Technology, Ningbo
University, Ningbo, Zhejiang 315211, China}

\begin{abstract}

Short gamma-ray bursts (GRBs) exhibiting a plateau phase provide
valuable insights into the post-merger activity of their central
engines. Although the physical origin of the plateau remains
uncertain, the magnetar energy injection model offers a compelling
explanation that reproduces the observed temporal and luminosity
features. However, previous studies relying solely on X-ray data
have suffered from strong parameter degeneracies when constraining
the magnetar parameters. Here we perform broadband afterglow
modeling on seven short GRBs with plateau features by combining
X-ray, optical, and radio observations within the framework of the
magnetar energy injection model. Key model parameters are derived
by using the Markov Chain Monte Carlo method. It is found that the
energy injection substantially modifies the afterglow dynamics in
most events. Compared with X-ray--only analyses, our broadband
modeling yields systematically a lower magnetic field strength and
a shorter spin period for the central magnetar, corresponding to a
higher injection luminosity. The study clearly shows that
incorporating multi-wavelength data effectively alleviates the
degeneracy between the magnetar parameters and X-ray radiative
efficiency. In addition, the distribution of our short GRBs
differs markedly from long GRBs when they are plotted on the
initial Lorentz factor versus gamma-ray energy plane. This offset,
consistent with the observed harder spectrum of short GRBs, may
serve as a useful diagnostic for investigating the progenitor as
larger samples are available.

\end{abstract}


\section{Introduction}
\label{sec:intro} Gamma-ray bursts (GRBs) are energetic transients
occurring at cosmological distances. They are widely believed to
come from internal dissipation processes in highly relativistic
outflows, such as internal shocks or magnetic reconnection
\citep{Piran:2004RvMP...76.1143P,Meszaros:2006RPPh...69.2259M,
Kumar:2015PhR...561....1K,zhang2018physics}. The prompt emission
of GRBs is typically non-thermal and the $\gamma$-ray spectrum can
empirically be well described by the so called Band function
\citep{Band:1993ApJ...413..281B}. Multi-wavelength spectral
analyses of GRB prompt phase support synchrotron radiation as the
dominant emission mechanism \citep{Oganesyan:2018A&A...616A.138O,
Oganesyan:2019A&A...628A..59O}. Based on their $T_{90}$ duration,
defined as the time interval during which the detector records
5$\%$ -- 95 $\%$ of the total observed photons, they are
conventionally divided into long GRBs ($T_{90}>2$ s) and short
GRBs ($T_{90}<2$ s) \citep{Kouveliotou:1993ApJ...413L.101K}. The
association of GRB 980425 with SN 1998bw established the link
between long GRBs and the collapse of massive stars
\citep{Galama:1998Natur.395..670G}, whereas the joint detection of
GW 170817 and GRB 170817A confirmed that binary neutron star
mergers can give rise to short GRBs
\citep{Abbott:2017ApJ...848L..12A,Abbott:2017ApJ...848L..13A}.
However, such a dichotomy is challenged by some special cases. For
instance, GRB 060614 had a long duration of $\sim 102$ s but
exhibited a negligible spectral lag, a typical feature of short
GRBs
\citep{Gehrels:2006Natur.444.1044G,Mangano:2007A&A...470..105M}.
Moreover, although it lies at a low redshift of $z=0.125$
\citep{DellaValle:2006Natur.444.1050D, Fynbo:2006Natur.444.1047F},
deep optical follow-up observations revealed no associated
supernova. Another enigmatic case is the short GRB 200826A, which
had a rest-frame duration of only $\sim$ 0.5 s, but exhibited a
supernova bump in its late optical and near-infrared afterglow
\citep{Zhang:2021NatAs...5..911Z,Rossi:2022ApJ...932....1R,Wang:2022ApJ...931L...2W}.
These anomalies have led to alternative classification schemes.
For example, researchers argued that GRBs could be grouped into
compact star events (Type I) and massive star events (Type II)
\citep{Zhang:2006Natur.444.1010Z,Kann:2011ApJ...734...96K}, which
incorporates multiple diagnostic criteria including the prompt
spectrum, empirical correlations (e.g., the Amati relation), host
galaxy characteristics, association with supernovae or kilonovae,
etc.

After the brief and highly variable prompt emission, a smoothly
evolving  afterglow emerges, whose multi-wavelength light curves
can typically be characterized by piecewise power-law segments,
with the spectrum well explained by the optically thin synchrotron
radiation model \citep{Sari:1998ApJ...497L..17S}. The \textit{Neil
Gehrels Swift Observatory} (hereafter \textit{Swift})
revolutionized early afterglow studies by virtue of its rapid
slewing capability, opening an observational window that had
previously been inaccessible \citep{Gehrels:2004ApJ...611.1005G}.
Based on the \textit{Swift} observations,
\cite{Zhang:2006ApJ...642..354Z} established a canonical X-ray
afterglow pattern comprising five components, among which the
plateau phase is particularly notable and is characterized by a
nearly constant or slowly decaying segment in the light curve.

Approximately half of long GRBs exhibit an X-ray plateau, which is
not expected in the standard external forward-shock model
\citep{Tang:2019ApJS..245....1T,Deng:2023ApJ...943..126D,Deng:2025Galax..13...15D}.
The shallow decay implies that the external shock may receive
additional energy injection from the central engine, leading to
deviations from the Blandford-McKee self-similar evolution
\citep{Blandford:1976PhFl...19.1130B}. In this context, the
millisecond magnetar model has attracted considerable attention,
as dipole radiation from a rotating magnetar can naturally produce
the observed plateau (\citealt{Dai:1998PhRvL..81.4301D,Zhang:2002ApJ...566..712Z,
Fan:2006MNRAS.372L..19F,Dall'Osso:2011A&A...526A.121D};
\citealt{Rowlinson:2014MNRAS.443.1779R,Rea:2015ApJ...813...92R};
\citealt{Tang:2019ApJS..245....1T,Zhang:2024ApJ...977..206Z}).
Observationally, the anti-correlation between the plateau duration
$T_{\rm a}$ and the corresponding X-ray luminosity $L_{\rm X}$
(the so-called Dainotti relation) is consistent with theoretical
predictions of the magnetar spin-down model
\citep{Dainotti:2008MNRAS.391L..79D,Dainotti:2010ApJ...722L.215D,
Dainotti:2013ApJ...774..157D,Stratta:2018ApJ...869..155S}.
Moreover, some GRBs exhibit a steep decay at the end of the
plateau, with slopes steeper than those predicted by high-latitude
emission, consistent with the collapse of a supra-massive magnetar
into a black hole
\citep{Chen:2017ApJ...849..119C,Lu:2018MNRAS.480.4402L}. While
many empirical correlations involving plateaus can be
satisfactorily interpreted in the magnetar framework, several
alternative explanations have also been proposed, including
off-axis structured jets \citep{Eichler:2006ApJ...641L...5E,
Beniamini:2020MNRAS.492.2847B,Beniamini:2022MNRAS.515..555B},
high-latitude emission from a structured jet
\citep{Oganesyan:2020ApJ...893...88O,Ascenzi:2020A&A...641A..61A,
Panaitescu:2025ApJ...989...33P}, the zero-point effect
\citep{Yamazaki:2009ApJ...690L.118Y,Ioka:2006A&A...458....7I,
Liang:2009ApJ...707..328L,Guidorzi:2025A&A...703A.101G},
mildly relativistic outflows propagating through a wind-like
environment \citep{Dereli-Begue:2022NatCo..13.5611D}, and
long-lived reverse shocks
\citep{Genet:2007MNRAS.381..732G,Hascoet:2014MNRAS.442...20H,Geng:2016ApJ...825..107G}.

A recent statistical analysis of X-ray afterglows from short GRBs
revealed that only about 18--37$\%$ of them exhibit a plateau
phase, a fraction markedly lower than that of long GRBs
\citep{Guglielmi:2024A&A...692A..73G,Tang:2019ApJS..245....1T}. If
such plateaus are indeed driven by magnetar spin-down, this
fraction places tight constraints on the probability of forming a
stable magnetar following binary neutron star mergers.
\cite{Guglielmi:2024A&A...692A..73G} further derived the magnetar
parameters based solely on the X-ray afterglow data, without
incorporating data of other wavelengths. Nevertheless, our
understanding of particle acceleration, magnetic field
amplification, and energy dissipation in relativistic shocks
remains incomplete
\citep{Sironi:2013ApJ...771...54S,Luongo:2021Galax...9...77L}.
Consequently, theoretical modeling generally relies on multiple
free parameters to describe the conversion of kinetic energy into
multi-wavelength afterglow emission, thereby introducing severe
parameter degeneracies \citep{Miceli:2022Galax..10...66M}.
Single-band data are therefore insufficient to place tight
constraints on the afterglow models
\citep{Garcia-Cifuentes:2024MNRAS.527.6752G}.

Multi-wavelength afterglow data play a pivotal role in alleviating
parameter degeneracies and probing the underlying physical
processes of GRBs
\citep{Laskar:2015ApJ...814....1L,Derishev:2021ApJ...923..135D}.
Observations across multiple bands provide complementary
diagnostics of the energetics, jet geometry, circumburst
environments, and shock microphysics. A prominent example is GRB
221009A, whose exceptionally rich afterglow data spanning from
radio to very-high-energy gamma-rays have provided an
unprecedented opportunity to test the underlying radiation
mechanisms such as synchrotron and synchrotron self-Compton (SSC)
processes
\citep{Bright:2023NatAs...7..986B,Laskar:2023ApJ...946L..23L,
LHAASOCollaboration:2023Sci...380.1390L,Ren:2024ApJ...962..115R,
Zheng:2024ApJ...966..141Z,Geng:2025arXiv250317765G}. It demonstrates
that single-band data are inadequate to fully constrain the physical
scenario, underscoring the necessity of integrating information across
multiple wavelengths.

In this study, we perform multi-band afterglow modeling on short
GRBs with plateau features in the framework of the magnetar energy
injection scenario. By combining multi-wavelength observational
data, we aim to test the magnetar-driven plateau interpretation
and constrain the magnetar parameters. This paper is organized as
follows. In Section \ref{sec2:modeling}, we outline the magnetar
energy injection framework and describe the multi-band afterglow
modeling approach. Section \ref{sec3} introduces the GRB sample
and details the fitting procedure. Section \ref{sec4} presents the
main fitting results and compares the derived magnetar parameters
with previous studies. In Section \ref{sec5}, we briefly
discuss the prompt--afterglow correlations and their implications.
Finally, we provide a brief summary and
discussion in Section \ref{sec:summary}.

\section{GRB Afterglow Modeling with Energy Injection}
\label{sec2:modeling}

\subsection{Shock Hydrodynamics}

After the prompt emission phase, the merged ejecta propagates into
the circumburst medium, driving an external shock that accelerates
electrons and produces multi-wavelength afterglow emission. As the
external shock sweeps up sufficient ambient material, it
transitions from a coasting phase to a self-similar deceleration
phase, during which the bulk Lorentz factor evolves as
$\Gamma(t)\propto t^{-3/8}$ \citep{Blandford:1976PhFl...19.1130B}.
In this study, we adopt the generic dynamical equation proposed by
\cite{Huang:1999MNRAS.309..513H} to calculate the temporal
evolution of the bulk Lorentz factor of the external shock. The
equation effectively describes the ultra-relativistic phase and
naturally reduces to the Sedov-Taylor solution in the
non-relativistic limit.

The plateau phase in the X-ray afterglow deviates from the decay
behavior predicted by the standard external shock model and
typically requires additional energy supply from the central
engine. Depending on the injection mechanism, the energy input can
be divided into two categories: one dominated by Poynting flux and
the other dominated by the kinetic energy of baryonic shells
\citep{Zhang:2002ApJ...566..712Z}.  In the former case, the
injected energy sustains the external shock, thereby delaying its
deceleration and leading to a contemporaneous rebrightening across
multiple wavelengths. In contrast, the latter scenario involves
collisions between successive shells, which generate forward and
reverse shocks propagating through the ejecta and contribute
additional emission components.

In this study, we consider the energy injection process dominated
by Poynting flux emitted by the central engine. The corresponding
dynamical equations governing the evolution of the external shock
are given by
\citep{Huang:1999MNRAS.309..513H,Kong:2010SCPMA..53S..94K,Liu:2010SCPMA..53S.262L,Geng:2013ApJ...779...28G}
\begin{align}
\frac{d\Gamma}{dm} & = -\frac{(\Gamma^2 - 1) - \frac{1 - \beta}{2 \beta c^3}
(1 - \cos \theta_j) L(t - R/c) \frac{dR}{dm}}{M_{\mathrm{ej}} + 2(1 - \epsilon)
\Gamma m + \epsilon m}, \\
\frac{dm}{dR} & = 2\pi R^2 (1 - \cos \theta_j) n m_p, \\
\frac{dR}{dt} & = \beta c \Gamma \left( \Gamma + \sqrt{\Gamma^2 - 1} \right).
\end{align}
Here, $m$ denotes the swept-up mass from the ambient medium, $c$ is the speed of
light, $M_{\rm ej}$ is the ejecta mass, $R$ is the radius of the external shock, and $t$
is the observer-frame time. The parameter $\epsilon$ denotes the radiative efficiency,
$\theta_j$ is the jet opening angle, and $\beta = \sqrt{1-1/\Gamma^2}$ represents the
dimensionless shock velocity. In addition, $m_{p}$ is the proton mass and $n$ is the
number density of the circumburst medium. The radial density profile $n(R)$ typically
follows one of two prescriptions: $n \propto R^0$ for a constant-density interstellar
medium (ISM) and $n \propto R^{-2}$ for a wind environment.

Continuous energy injection with a luminosity $L$ into the
external shock can produce afterglow emission exhibiting a plateau
feature. Magnetars are widely regarded as promising candidates for
the central engine of GRBs, and several empirical correlations
among plateau parameters support a magnetar-driven origin for this
phase (e.g.,
\citealt{Dainotti:2008MNRAS.391L..79D,Tang:2019ApJS..245....1T,
Deng:2023ApJ...943..126D,Deng:2025Galax..13...15D}). The temporal
evolution of the Poynting flux luminosity, derived from the
decreasing rotational energy of a newly formed magnetar, can be
expressed as
\begin{eqnarray}
L(t) & = & L_0 \left(1 + \frac{t}{t_{0,\mathrm{em}}} \right)^{-2},
\end{eqnarray}
where $L_0$ denotes the initial magnetic dipole luminosity of the
magnetar. It can be written as
\begin{eqnarray}
L_0  =  \frac{I \Omega_0^2}{2 t_{0,\mathrm{em}}}  = \frac{B_p^2
R_{\mathrm{NS}}^6 \Omega_0^4}{6 c^3} \simeq 1.0 \times
10^{47}~\mathrm{erg~s^{-1}}\, \left(\frac{B_{p}}{10^{14}~\mathrm
G} \right)^2 \left(\frac{P_{0}}{10^{-3}~\mathrm s} \right)^{-4}
\left(\frac{R_{\mathrm{NS}}}{10^{6}~\mathrm{cm}} \right)^6,
\end{eqnarray}
where $I$, $B_p$, $P_0$, and $\Omega_0$ denote the moment of
inertia, the magnetic field strength at the polar cap, the initial
spin period, and the initial angular velocity of the magnetar,
respectively. The quantity $t_{0,\mathrm{em}}$ is the
characteristic spin-down timescale due to magnetic dipole
radiation, i.e.
\begin{equation}
t_{0,\mathrm{em}} = \frac{3c^3 I}{B_p^2 R^6 \Omega_0^2} \simeq 2.1
\times 10^5~ \left(\frac{I}{10^{45} \mathrm{g~cm^{2}}}\right)
\left(\frac{B_p}{10^{14}~\mathrm{G}}\right)^{-2}
\left(\frac{P_0}{10^{-3}~\mathrm{s}}\right)^{2}
\left(\frac{R_{\mathrm
NS}}{10^{6}~\mathrm{cm}}\right)^{-6}~\mathrm{s}.
\end{equation}

\subsection{Afterglow Emission}
\label{afterglow_emission}
As the external shock propagates through the circumburst medium, its kinetic energy
is converted into the internal energy of particles, which subsequently radiate across
multiple wavelengths via synchrotron emission and SSC process. Numerical simulations
suggest that the energy spectrum of shock-accelerated electrons does not strictly follow
a power-law distribution and may contain a population of low-energy thermal electrons
\citep{Spitkovsky:2008ApJ...673L..39S,Gao:2024ApJ...971...81G}. For simplicity, we assume
that the injected electron population follows a power-law distribution with a spectral
index $p$, i.e., $dN_e/d\gamma_e \propto \gamma_e^{-p}$. The minimum Lorentz factor
$\gamma_{\rm e,min}$ of electrons accelerated by the external shock is
\citep{Huang:2000ApJ...543...90H}
\begin{eqnarray}
\gamma_{e,\min}' & = & \xi_e (\Gamma - 1) \frac{m_p (p - 2)}{m_e (p - 1)} + 1,
\end{eqnarray}
where $\xi_e$ represents the fraction of proton energy transferred to the
shock-accelerated electrons. Here, variables with a prime ($'$) superscript are defined
in the comoving frame of the shock. These electrons lose energy through synchrotron
radiation and SSC cooling, resulting in the time-dependent evolution of their energy
distribution. Accordingly, it is convenient to define the cooling Lorentz factor
\citep{Sari:1998ApJ...497L..17S},
\begin{eqnarray}
\gamma_{e,c}' & = & \frac{6 \pi m_e c}{\sigma_{\mathrm{T}} B'^2 t'(1 + Y)},
\end{eqnarray}
where $Y$ denotes the ratio of SSC power to synchrotron power and $B'$ represents the
magnetic field strength in the shock frame \citep{Huang:2000ApJ...543...90H},
\begin{eqnarray}
\frac{B'^2}{8\pi} & = & \xi_B^2 \frac{\hat{\gamma} \Gamma + 1}{\hat{\gamma} - 1}
(\Gamma - 1) n m_p c^2,
\end{eqnarray}
with $\xi_B^2$ and $\hat{\gamma}=\frac{4\Gamma+1}{3\Gamma}$ being the magnetic energy
equipartition parameter and adiabatic index, respectively.

In the standard synchrotron radiation model, the emission spectrum of shock-accelerated
electrons follows a multi-segment power-law function. The characteristic frequencies
$\nu_m' = \frac{3}{4\pi} \frac{\gamma_{\rm e,min}'^2eB'}{m_ec} $ and
$\nu_c' = \frac{3}{4\pi}\frac{\gamma_{\rm e,c}'^2eB'}{m_ec}$ serve as the spectral break
frequencies. For the fast-cooling regime ($\nu_c' < \nu_{\min}'$), the flux density at
frequency $\nu'$ is given by \citep{Sari:1998ApJ...497L..17S}
\begin{eqnarray}
L_\nu'(\nu') & = & L_{\nu,\max}' \times
\begin{cases}
\left( \dfrac{\nu'}{\nu_c'} \right)^{1/3}, & \nu' < \nu_c', \\[10pt]
\left( \dfrac{\nu'}{\nu_c'} \right)^{-1/2}, & \nu_c' < \nu' < \nu_m', \\[10pt]
\left( \dfrac{\nu'}{\nu_m'} \right)^{-p/2} \left( \dfrac{\nu_m'}{\nu_c'} \right)^{-1/2},
& \nu_m' < \nu < \nu_{\max}'.
\end{cases}
\end{eqnarray}
Here, $\nu'_{\max}$ denotes the characteristic frequency corresponding to the maximum Lorentz
factor, $\gamma'_{\rm{e,max}} \simeq 10^8 (B'/1~\rm{G})^{-1/2}$, of shock-accelerated
electrons. The quantity $L'_{\nu,\max} = N_e \frac{\sqrt{3} e^3 B'}{m_e c^2}$ represents the
specific luminosity in the comoving frame of the shock, where $N_e$ is the number of
electrons contributing to the radiation. For the slow-cooling scenario
($\nu_c' > \nu_{\min}'$), one can obtain
\begin{eqnarray}
L_\nu'(\nu') & = & L_{\nu,\max}' \times
\begin{cases}
\left( \dfrac{\nu'}{\nu_m'} \right)^{1/3}, & \nu' < \nu_m', \\[10pt]
\left( \dfrac{\nu'}{\nu_m'} \right)^{-(p-1)/2}, & \nu_m' < \nu' < \nu_c', \\[10pt]
\left( \dfrac{\nu'}{\nu_c'} \right)^{-p/2} \left( \dfrac{\nu_c'}{\nu_m'} \right)^{-(p-1)/2},
& \nu_c' < \nu' < \nu_{\max}'.
\end{cases}
\end{eqnarray}

In the observer frame, the flux density at frequency $\nu$ is given by
\citep{Geng:2018ApJS..234....3G,Ren:2024ApJ...962..115R}
\begin{eqnarray}
\mathcal{F}_\nu(\nu) & = & \frac{1 + z}{4\pi D_L^2} \int_0^{\theta_j}
L_\nu'((1+z)\nu/\mathcal{D})\, \mathcal{D}^3 \, \frac{\sin\theta}{2} \, d\theta,
\end{eqnarray}
where $\mathcal{D} = \frac{1}{\Gamma(1-\beta \cos \theta)}$
denotes the Doppler factor, and $z$ is the redshift of the source.
In this study, we adopt a flat $\Lambda$CDM cosmological model
with $H_0 = 67.4~\mathrm{km~s^{-1}~Mpc^{-1}}$, $\Omega_\mathrm{m}
= 0.315$, and $\Omega_\Lambda = 0.685$ to calculate the luminosity
distance $D_{\rm L}$
\citep{PlanckCollaboration:2020A&A...641A...6P}. The effect of the
equal-arrival-time surface is considered, which is defined by the
condition that photons emitted at different radii and angles
arrive at the observer simultaneously. This constraint is
determined by \citep{Huang:2000ApJ...543...90H,Geng:2016ApJ...825..107G}
\begin{eqnarray}
t_{\rm obs} & = & (1+z) \int \frac{1 - \beta \cos \theta}{\beta c} \, dr \equiv \text{const.}
\end{eqnarray}
In addition, synchrotron self-absorption is incorporated into the
afterglow modeling to more accurately reproduce the temporal
evolution of the radio afterglow
\citep{Wu:2003MNRAS.342.1131W,Geng:2016ApJ...825..107G}.

\section{Sample Selection and Fitting Strategy}
\label{sec3}

Compared to long GRBs, the afterglows of short GRBs are generally
fainter \citep{Berger:2014ARA&A..52...43B,Fong:2015ApJ...815..102F},
which makes it challenging to identify plateau phases in these events.
\cite{Guglielmi:2024A&A...692A..73G} compiled a sample of 85 short
GRBs with measured redshifts, all of which were detected by the
\textit{Swift} satellite between May 2005 and December 2021. After
excluding low signal-to-noise events, \cite{Guglielmi:2024A&A...692A..73G}
fitted the X-ray afterglow light curves of the remaining GRBs using
a smoothly broken power-law function. Utilizing the magnetar energy
injection model proposed by \cite{Dall'Osso:2011A&A...526A.121D},
they estimated the magnetic field strengths and initial spin periods
of the central magnetars for 13 short GRBs exhibiting plateau features.

\begin{figure}[ht!]
\gridline{\fig{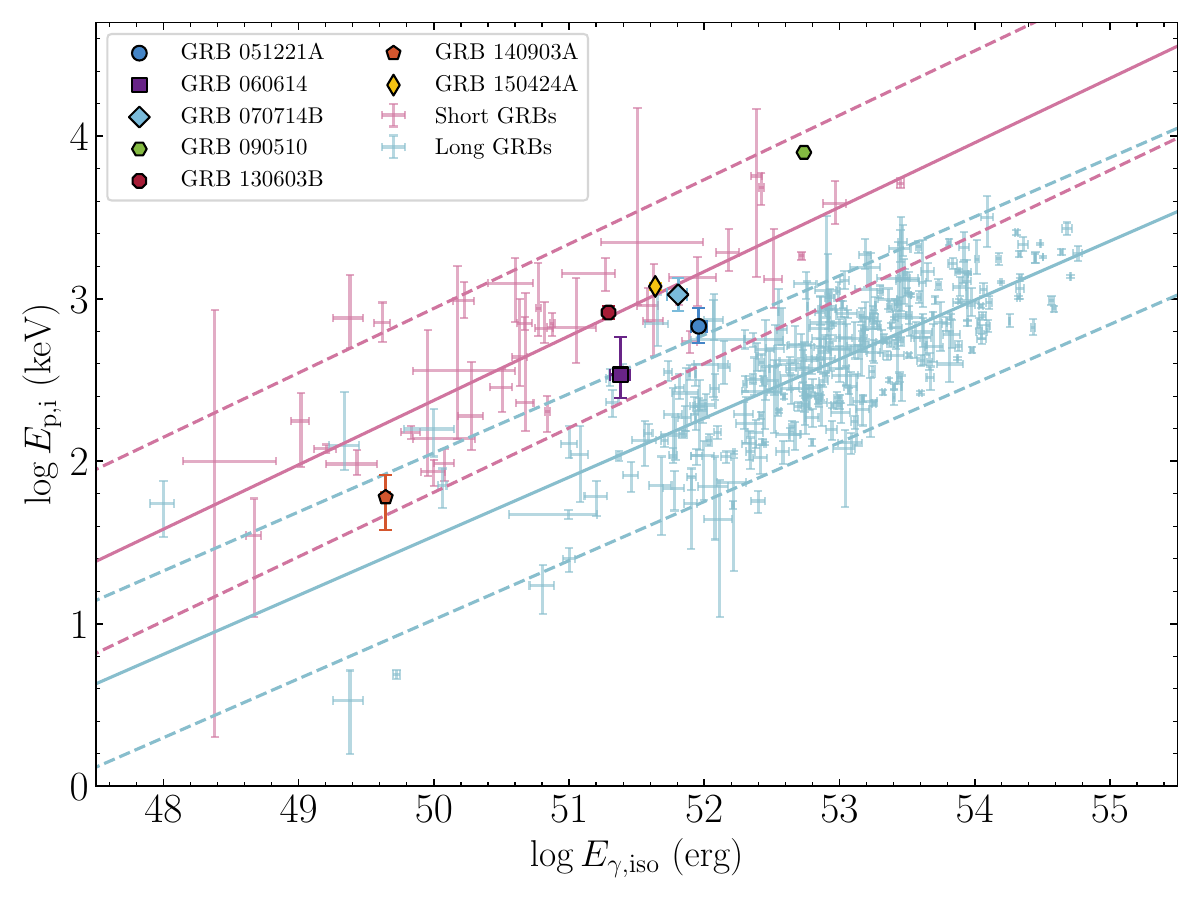}{0.6\textwidth}{}} \caption{Correlation
between the rest-frame spectral peak energy $E_{\mathrm{p,i}}$ and
the isotropic-equivalent energy $E_{\gamma,\mathrm{iso}}$ (the
Amati relation). Our sample is compared with long GRBs (cyan
points) and short GRBs (pink points), of which the data are taken
from \cite{Minaev:2020MNRAS.492.1919M}. The solid and dashed lines
represent the best-fit relations and their $2\sigma$ regions.
\label{fig1}}
\end{figure}

We applied a more stringent criterion on the magnetar sample of
\cite{Guglielmi:2024A&A...692A..73G} by selecting events with
available multi-wavelength afterglow observations. Specifically,
seven short bursts -- GRBs 051221A, 060614, 070714B, 090510,
130603B, 140903A, and 150424A -- are included for broadband
modeling within the framework of the magnetar energy injection
model as described in Section~\ref{sec2:modeling}.
Figure \ref{fig1} shows the distribution of our short GRB sample
in the $E_{\mathrm{p,i}}-E_{\gamma,\mathrm{iso}}$ (Amati relation)
plane. These bursts lie systematically above the best-fit relation
for long GRBs, consistent with the typical offset observed for
short GRBs. It suggests that these events are more likely to
originate from compact object mergers rather than collapsars. For
simplicity, we adopt a uniform ISM environment instead of a
wind-like density profile in the afterglow modeling.

In the framework of the energy injection process, nine free
parameters are introduced in total, i.e. $E_{\mathrm{K,iso}}$,
$\Gamma$, $n$, $\theta_j$, $p$, $\xi_e$, $\xi_B^2$, $B_p$, and
$P_0$. The observer is assumed to be on the jet axis with a zero
viewing angle ($\theta_v = 0$). Moreover, given that most data
points in our sample are within $10^6$~s after the GRB trigger,
lateral expansion of the jet is not expected to significantly
affect the light curve
\citep{Huang:2000MNRAS.316..943H,Huang:2000ApJ...543...90H,Xu:2023A&A...679A.103X}.
Consequently, this effect is neglected to reduce model complexity.

The Markov Chain Monte Carlo (MCMC) method, implemented with the \texttt{emcee} package
\citep{Foreman-Mackey:2013PASP..125..306F}, is employed to jointly fit the multi-wavelength
observational data and to obtain the posterior probability distributions of the model
parameters. The logarithmic likelihood function adopted in the fitting process is given
by \citep{Zhang:2024ApJ...972..170Z}
\begin{eqnarray}
\log \mathcal{L} & = & -\frac{1}{2} \sum_{i} \left\{ \left( \frac{y_i - \hat{y}_i}{\sigma_{y_i}}
\right)^2 + \ln \left( 2\pi\sigma_{y_i}^2 \right) \right\},
\end{eqnarray}
where $y_i$ denotes the observed logarithmic flux density of the $i$-th afterglow data point,
$\sigma_{y_i}$ is its corresponding observational uncertainty, and $\hat{y}_i$ represents the
model-predicted value. All observational data points are weighted by their respective
uncertainties during the fitting process to ensure unbiased parameter estimation.

\section{Numerical results}
\label{sec4}

We have collected multi-band afterglow data of the seven short
GRBs introduced in Section \ref{sec3}. In this section, we present
the modeling results for their broadband afterglows. The X-ray
afterglow data were downloaded from the \textit{Swift}--XRT light
curve repository
\footnote{\url{https://www.swift.ac.uk/xrt_curves/}}
\citep{Evans:2007A&A...469..379E,Evans:2009MNRAS.397.1177E} and
converted into flux densities at 1 keV. The optical light curves
were obtained from the optical GRB catalogue
\footnote{\url{https://grblc-catalog.streamlit.app/}} compiled by
\cite{Dainotti:2024MNRAS.533.4023D}, which provides data already
corrected for Galactic extinction. Among the seven bursts, four
have available radio afterglow observations, i.e. GRB 051221A
\citep{Soderberg:2006ApJ...650..261S}, GRB 130603B
\citep{Fong:2014ApJ...780..118F}, GRB 140903A
\citep{Troja:2016ApJ...827..102T}, and GRB 150424A
\citep{Fong:2015GCN.17804....1F}. We thus have collected the radio
afterglow data from the corresponding literature. Using the
magnetar energy injection model described in Section
\ref{sec2:modeling}, we model their multi-wavelength afterglows
and compare the theoretical results with observations. Figure
\ref{fig2} displays the best-fit light curves across multiple
bands for each burst, while Figure \ref{fig3} presents the
corresponding posterior distributions of the model parameters
derived from the MCMC method, which are also summarized in Table
\ref{table1}. As illustrated in Figures \ref{fig2}--\ref{fig3},
the magnetar energy injection model successfully reproduces the
multi-band afterglow evolution of these short GRBs. For instance,
in GRB 140903A, as the external shock is continuously refreshed by
energy injection from the central magnetar, the X-ray and optical
light curves deviate from the simple power-law decay expected by
the standard forward shock model after $10^3$ s, showing a
distinct flattening trend.

Our fitting results indicate that the initial Lorentz factors of
the jets all exceed 100, confirming that these events correspond
to successful bursts rather than baryon-rich ``dirty'' fireballs
\citep{Huang:2002MNRAS.332..735H}. The inferred circumburst densities
cluster around $10^{-3}$ -- $10^{-1}\mathrm{cm^{-3}}$, consistent
with the density range reported for the short GRB population
\citep{Fong:2015ApJ...815..102F}, and are indicative of a compact
binary-merger origin. Moreover, the jet opening angles derived
from our broadband modeling are in line with the analytical
estimates based on the jet-break times, reinforcing the robustness
of our results.

Based on the best-fit parameters, the magnetic field strengths of
the central magnetars in our short GRB sample are $2.69\times10^{14}$ G (GRB 051221A),
$8.13\times10^{14}$ G (GRB 060614), $9.33\times10^{14}$ G (GRB
070714B), $3.98\times10^{12}$ G (GRB 090510), $1.26\times10^{15}$ G
(GRB 130603B), $6.61\times10^{14}$ G (GRB 140903A), and
$5.62\times10^{14}$ G (GRB 150424A), respectively, with corresponding
initial spin periods of $1.10$, $1.91$, $1.82$, $4.57$, $2.95$, $1.07$,
and $1.74$ ms. The injected energy is estimated as $E_{\rm inj}\simeq L_0
T_{\rm em}$, where $T_{\rm em}=\min(t_{0,\rm em},t_{\rm
break}/(1+z))$, with $t_{\rm break}$ denotes the ending time of
the plateau as reported in \citet{Guglielmi:2024A&A...692A..73G}.
To quantify the relative significance of energy injection,
we define the dimensionless parameter
$\zeta = E_{\rm inj}/E_{\rm K,iso}$. The resulting
$\zeta$ values for our sample are approximately 1.59, 174.67,
0.33, $2.2\times10^{-8}$, 0.035, 5.65, and 6.20, respectively.
Except for GRB 090510, the inferred $\zeta$ values
reveal that magnetar energy injection plays a non-negligible role
in shaping the afterglow dynamics. GRB 090510, however, with
$\zeta\sim\times10^{-8}$, shows little evidence of additional
energy input, implying that its afterglow is governed primarily by
the initial kinetic energy, consistent with the standard external
shock scenario.

Fermi-LAT detected photons up to $\sim$30 GeV from GRB 090510,
implying a lower limit of $\Gamma_0 \gtrsim 1200$ for the jet
Lorentz factor, as inferred from $\gamma$-$\gamma$ opacity
constraints \citep{Ackermann:2010ApJ...716.1178A}. The
deceleration timescale $t_{\rm dec}$ of the external shock
is highly sensitive to the initial Lorentz factor
\citep{Meszaros:2006RPPh...69.2259M,Geng:2025ApJ...984L..65G}:
\begin{equation}\label{func:t_dec}
t_{\mathrm{dec}} \simeq 189 ~(1+z)
\left( \frac{E_{\mathrm{K,iso}}}{10^{53}\ \mathrm{erg}} \right)^{1/3}
\left( \frac{n}{1~\mathrm{cm^{-3}}} \right)^{-1/3}
\left( \frac{\Gamma_0}{100} \right)^{-8/3} \mathrm{\ s}.
\end{equation}
Our modeling yields $t_{\rm dec}\approx0.25~\rm s$ for GRB 090510.
\cite{Ghirlanda:2010A&A...510L...7G} interpreted the onset of the
GeV emission as the jet deceleration epoch and derived
$\Gamma_0\sim2000$, in agreement with our modeling results. This
finding further supports that the GeV emission of GRB 090510
arises during the afterglow phase. In contrast,
\cite{Zhang:2025ApJ...991..209Z} derived $\Gamma_0 \approx 72$ by
adopting the optical afterglow peak time ($\sim$1500 s) as the
deceleration epoch. Since $\Gamma_0 \propto t_{\rm dec}^{-3/8}$, a
later peak time naturally results in a smaller $\Gamma_0$.
However, this optical peak is more plausibly associated with the
passage of the characteristic synchrotron frequency $\nu_m$
through the observed band, rather than with the dynamical
deceleration of the outflow. \cite{Liang:2013ApJ...774...13L}
modeled the optical afterglow using a smoothly broken power-law
function and obtained a rising slope of $0.47\pm0.14$. During the
Blandford-McKee self-similar deceleration phase, this slope
agrees with the theoretical expectation of $F_\nu\propto
\nu_m^{-1/3}\propto t^{0.5}$ for $\nu<\nu_m$ in the slow-cooling
regime. The timescale when the characteristic frequency $\nu_m$
crosses the observing frequency $\nu$ is given by
\citep{Sari:1998ApJ...497L..17S,Geng:2016ApJ...825..107G}
\begin{equation}
t_{\mathrm m} \simeq  10^4 ~
\left( 1+z \right)^{1/3}
\left( \frac{p-2}{p-1} \right)^{4/3}
\left( \frac{\xi_e}{0.1} \right)^{4/3}
\left( \frac{\xi_B^2}{0.1} \right)^{1/3}
\left( \frac{E_{\mathrm{K,iso}}}{10^{53}\ \mathrm{erg}} \right)^{1/3}
\left( \frac{\nu}{10^{15}~\mathrm{Hz}} \right)^{-2/3} \mathrm{s}.
\end{equation}
Substituting the best-fit parameters and
$\nu=8.8\times10^{14}~\rm{Hz}$ for the \textit{Swift}-UVOT U band
yields $t_{\mathrm{m}} \approx 1.42\times10^3 \rm{s}$, which is in
good agreement with our numerical modeling.

In Figure \ref{fig4}, we compare the magnetar parameter
distributions obtained in this work with those reported by
\cite{Guglielmi:2024A&A...692A..73G}, who only used X-ray data in
their modeling. For reference, we also include the short
GRB sample of \cite{Stratta:2018ApJ...869..155S}, which overlaps
with our sample in four events, namely GRBs~051221A, 060614,
070714B, and 090510. Both \cite{Stratta:2018ApJ...869..155S} and
\cite{Guglielmi:2024A&A...692A..73G} derived the magnetar
parameters within the energy injection framework of
\cite{Dall'Osso:2011A&A...526A.121D}, but they adopted different
spin-down prescriptions: \cite{Guglielmi:2024A&A...692A..73G}
assumed an ideal magnetic dipole scenario, whereas
\cite{Stratta:2018ApJ...869..155S} included magnetospheric
resistivity. This difference may account for their modest offset
in the $B_{\rm p}$--$P_0$ plane, while the overall distributions
remain largely consistent.

In the $B_{\rm p}$--$P_0$ plane, our results cluster
toward the region with a relatively lower magnetic field strength
and a shorter spin period, whereas the X-ray-only samples of
\citet{Guglielmi:2024A&A...692A..73G} and
\citet{Stratta:2018ApJ...869..155S} favor a stronger magnetic
field and a longer spin period. All samples yield
$t_{0,\mathrm{em}}$ values in the range of $10^3$--$10^5$ s,
indicating broadly similar plateau durations. However, reproducing
the shallow decay observed in both X-ray and optical bands
requires a higher initial energy injection in our fits,
corresponding to a stronger magnetic dipole luminosity. As
demonstrated by \cite{Zhang:2002ApJ...566..712Z}, once the
injected energy approaches or surpasses the initial kinetic energy
of the fireball, the external shock dynamics is substantially
modified, producing plateau or bump features. Consequently, for a
given $t_{0,\mathrm{em}}$, a lower $B_{\rm p}$ combined with a
shorter $P_0$ naturally implies a higher $L_0$, in agreement with
our broadband modeling. The difference between our results
and the X-ray-only analyses in the $B_{\rm p}$--$P_0$ plane mainly
reflects the assumption of isotropic energy injection and the
usage of multi-wavelength afterglow constraints. Assuming an
isotropic energy injection naturally demands a larger energy
budget than the beaming-corrected treatments used by
\cite{Stratta:2018ApJ...869..155S} and
\cite{Guglielmi:2024A&A...692A..73G}, shifting the inferred
magnetar parameters along the spin-up line in the $B_{\rm
p}$--$P_0$ plane \citep{Stratta:2018ApJ...869..155S}. Moreover,
low-frequency afterglow data (e.g., optical and radio) offer
critical constraints on the external shock energy and circumburst
density. A joint multi-band analysis effectively alleviates the
degeneracies among magnetic field strength, spin period, and
radiation efficiency, thereby yielding more reliable estimates of
the injection power.

Figure \ref{fig5} depicts the correlation between $E_{\mathrm{K,iso}}$ and
$E_{\gamma,\mathrm{iso}}$ in our short GRB sample. The prompt radiation
efficiency $\eta_\gamma$ is defined in terms of $E_{\mathrm{K,iso}}$ and
$E_{\gamma,\mathrm{iso}}$ as
\begin{equation}
\eta_{\gamma} = \frac{E_{\gamma,\mathrm{iso}}}{E_{\gamma,\mathrm{iso}} + E_{\mathrm{K,iso}}}.
\end{equation}
GRB 051221A and GRB 130603B exhibit relatively low efficiencies
($\eta_\gamma \lesssim 0.1$), in line with expectations for
baryon-dominated jets. By contrast, GRB 060614 displays an
extremely high efficiency ($\eta_\gamma \gtrsim 0.95$). It may
result from the long-lived plateau phase and/or the absence of
early afterglow observations, both of which bias the fit toward a
lower initial kinetic energy, consequently amplifying the
contribution portion of energy injection and inflating the
efficiency. Nevertheless, such a high-efficiency burst likely
involves a highly magnetized outflow, in agreement with the
internal-collision-induced magnetic reconnection and turbulence
(ICMART) model proposed by \cite{Zhang:2011ApJ...726...90Z}.
Overall, the wide dispersion of $\eta_\gamma$ underscores the
intrinsic diversity of jet compositions and dissipation mechanisms
among short GRBs, suggesting that their outflows are likely hybrid
in nature, including both magnetized and baryonic components.

A clear correlation has been reported between $\Gamma_0$ and
$E_{\gamma,\rm{iso}}$ in previous studies
\citep{Liang:2010ApJ...725.2209L,Liang:2013ApJ...774...13L,
Ghirlanda2012MNRAS.420..483G,Ghirlanda:2018A&A...609A.112G}.
The $\Gamma_0$--$E_{\gamma,\rm{iso}}$ correlation obtained from our
short GRB sample is compared with previous results of long GRBs
in Figure \ref{fig6}. We see that short GRBs lie systematically
above the long GRB region, exhibiting higher $\Gamma_0$ values
for a given $E_{\gamma,\rm{iso}}$. This offset is consistent with
the observational result that short GRBs generally have a harder
prompt spectrum than long ones. Non-parametric rank tests yield
a Spearman coefficient of $\rho=0.75$ ($p=0.05$) and a Kendall
coefficient of $\tau=0.62$ ($p=0.07$) for our sample. In addition,
the Pearson correlation coefficient is $r=0.49$ ($p=0.26$).
The correlation is not statistically significant, which is likely
due to the small sample size and the influence of outliers.
Although the limited statistics prevent a definitive determination
of the $\Gamma_0$--$E_{\gamma,\rm{iso}}$ correlation for short GRBs,
the overall monotonic trend suggests that a more energetic short burst
tends to possess a larger initial Lorentz factor. The systematic
offset between short and long GRBs in the
$\Gamma_0$--$E_{\gamma,\rm{iso}}$ plane is reminiscent of their
distinct distributions in the Amati relation. It implies that the
$\Gamma_0$--$E_{\gamma,\rm{iso}}$ correlation could serve as a
valuable complement to the Amati relation, providing an additional
empirical diagnostic for distinguishing progenitor types and
probing the energy release mechanisms of GRBs.

\section{Prompt-afterglow correlations and implications}
\label{sec5} 

From a phenomenological standpoint, GRBs with an afterglow
plateau exhibit a pronounced anti-correlation between the
luminosity at the end of the plateau phase and its duration,
commonly referred to as the Dainotti relation
\citep{Dainotti:2008MNRAS.391L..79D,
Dainotti:2010ApJ...722L.215D}. This relation remains statistically
significant after correcting for redshift evolution and selection
effects \citep{Dainotti:2013MNRAS.436...82D}, implying that it is
not driven by observational biases and may reflect physical
constraints shared among GRB classes. The inclusion of additional
observables allows the Dainotti relation to be extended to
three-parameter forms, such as the Dainotti fundamental plane
(also referred to as the 3D Dainotti relation)
\citep{Dainotti:2016ApJ...825L..20D,Dainotti:2017ApJ...848...88D}
and the $L_{X}-T_{a}-E_{\gamma,\rm iso}$ correlation
\citep{Xu:2012A&A...538A.134X,Tang:2019ApJS..245....1T}, which
typically exhibit a smaller intrinsic scatter than the original
two-parameter relation. Similar plateau correlations have also
been reported in the optical and radio bands
\citep{Dainotti:2020ApJ...905L..26D,
Levine:2022ApJ...925...15L,Dainotti:2022ApJS..261...25D},
providing further support for the interpretation that the GRB
external shock is substantially modified by sustained energy
injection from the central engine \citep{Li:2026arXiv260101586L}.

Typically, short GRBs are characterized by a lower
isotropic-equivalent prompt emission energy and kinetic energy
than long GRBs. For our short GRB sample, multi-wavelength
afterglow modeling constrains the initial spin period of the
central magnetars to a few milliseconds, approaching the Keplerian
limit for neutron stars \citep{Lattimer:2004Sci...304..536L}. By
contrast, the high-luminosity plateau observed in a subset of long
GRBs requires a substantially stronger energy injection. Since the
rotational energy of a newborn millisecond magnetar is limited to
a few $10^{52}$ erg, the magnetar scenario may be energetically
challenged when applied to the events with the most luminous
plateau. \cite{Lenart:2025JHEAp..4700384L} recently proposed an
alternative scenario in which the plateau emission is powered by
spin-down energy extracted from a rotating black hole surrounded
by a magnetically arrested disk. Under reasonable assumptions,
this model naturally produces an anti-correlation between the
plateau luminosity and its end time, with a power-law slope of
$-1$, in good agreement with the observed Dainotti relation. These
results indicate that the origin of the Dainotti relation may
involve multiple mechanisms.

Empirical correlations among GRB observables are not only
valuable for cosmological applications
\citep{Amati:2006MNRAS.372..233A,
Hu2021MNRAS.507..730H,Cao:2022MNRAS.516.1386C,Dainotti:2022MNRAS.514.1828D,
Wang:2022ApJ...924...97W,Cao:2025JCAP...09..081C}, but also
provide useful clues on their origins
\citep{Yang:2022Natur.612..232Y,Ierardi:2025arXiv251016108I}.
\cite{Minaev:2020MNRAS.492.1919M} introduced an energy-hardness
(EH) parameter based on the Amati relation to separate
collapsar-origin bursts from merger-origin events, yielding a more
robust classification than duration-based criteria.
\cite{Lenart:2025JHEAp..4700384L} further defined a plateau shift
(PS) parameter from the Dainotti fundamental plane and constructed
the EH--PS diagram, which provides joint constraints from prompt
and afterglow properties on GRB origins. Within this framework,
short GRBs form a distinct population and tend to lie
systematically below the Dainotti fundamental plane, implying that
their plateaus are generally underluminous compared with those of
long GRBs \citep{Dainotti:2017ApJ...848...88D}. Consistently, in
the 2D Dainotti relation they preferentially occupy the
lower-luminosity regime at comparable plateau durations
\citep{Dainotti:2013ApJ...774..157D}. Such offsets may reflect the
difference in underlying central engine properties of short and
long GRBs, but could also be influenced by the environmental
conditions that affect how efficiently the injected energy is
converted into the afterglow emission. Together with the
$\Gamma_0$--$E_{\gamma,\rm{iso}}$ correlation discussed in Section
\ref{sec4}, these empirical relations provide complementary
constraints on GRB origins.

\section{Conclusions and discussion}
\label{sec:summary}

In this study, we concentrate on short GRBs with a plateau phase
in the afterglow stage. A total number of seven events are
included in our sample, for which multi-band afterglow light
curves are available. Adopting the magnetar energy injection
model, we perform joint multi-band afterglow fits using the MCMC
method and obtain posterior distributions of the model parameters.
It is found that the magnetar energy injection scenario
successfully reproduces the overall afterglow evolution across the
sample. For GRB 090510, however, the total injected energy is
negligible as compared with the initial kinetic energy of the
fireball. By comparing the isotropic-equivalent prompt
$\gamma$-ray energy release with the initial kinetic energy of the
fireball, the radiative efficiency is derived for each burst. The
results further suggest that, beyond conventional baryon-dominated
fireballs, the high efficiency hinted in several GRBs is more
likely due to a highly magnetized outflow.

For GRBs exhibiting an X-ray plateau, the magnetic field strength
and initial spin period of the central magnetar can be derived in
the framework of the magnetar energy injection model by
considering the dipolar spin-down process. Although this approach
is straightforward, it possesses inherent limitations. Broadband
afterglow observations provide tighter constraints on the kinetic
energy of the fireball and the circumburst density
\citep{Fong:2015ApJ...815..102F}. Analyses relying solely on X-ray
data can introduce systematic biases in parameter estimation. In
our analysis, the inclusion of multi-wavelength data produces a
noticeably different distribution of magnetar parameters in the
$B_{\rm p}-P_0$ plane as compared with the results of
\cite{Guglielmi:2024A&A...692A..73G} and
\cite{Stratta:2018ApJ...869..155S}. We note that the difference in
the assumed beaming angle of the energy injection may contribute
to the observed offset, leading the inferred parameters to shift
along the spin-up line in the $B_{\rm p}$--$P_0$ plane. Moreover,
the assumptionthat the plateau luminosity scales directly with the
magnetic dipole radiation power leads to pronounced degeneracy
among the X-ray radiative efficiency, magnetic field strength, and
initial spin period. For a particular plateau luminosity and
duration, a higher radiative efficiency tends to result in a
stronger magnetic field and a longer initial spin period, leading
to a lower intrinsic dipole luminosity. Integrating
multi-wavelength afterglow data therefore plays a crucial role in
alleviating such degeneracies and improving the parameter
estimation.

For long GRBs displaying plateau features in both X-ray and
optical bands, \cite{Ronchini:2023A&A...675A.117R} demonstrated
that most events can be reproduced by a single-zone synchrotron
emission model. This result is consistent with the
Poynting-flux-dominated energy injection scenario, in which the
injected energy does not generate new emission regions but instead
delays the deceleration of the external shock, thereby reproducing
the observed plateau. Consequently, the total injected energy
should be comparable to the fireball's initial kinetic energy in
order to produce an apparent dynamical effect
\citep{Zhang:2002ApJ...566..712Z}. Class-dependent
differences in the plateau luminosity, with long GRBs generally
exhibiting a brighter plateau, have been well established through
studies of the 2D and 3D Dainotti relations
\citep{Dainotti:2013ApJ...774..157D,Dainotti:2017ApJ...848...88D,
Dainotti:2017A&A...600A..98D}. Using an updated sample,
\cite{Lenart:2025JHEAp..4700384L} further investigated the
Dainotti fundamental plane and found that short GRBs
systematically lie below the plane, indicating an intrinsically
fainter plateau than long GRBs. In addition, large-sample studies
have shown that the energetics of short GRBs is systematically
lower than that of long ones
\citep{Amati:2006MNRAS.372..233A,Ghirlanda:2009A&A...496..585G,
Tsutsui:2013MNRAS.431.1398T,Dainotti:2018PASP..130e1001D,Tang:2019ApJS..245....1T}.
Since the rotational energy of a neutron star is limited, the
magnetar model is difficult to account for the exceptionally
luminous plateaus observed in some long GRBs. By contrast, energy
extracted from a Kerr black hole via the Blandford-Znajek (BZ)
mechanism \citep{Blandford:1977MNRAS.179..433B}, where the outflow
is collimated into a Poynting-flux dominated jet, offers a more
plausible explanation for such luminous plateaus
\citep{Li:2018ApJS..236...26L}. In particular, within the
magnetically arrested disk state, such a scenario naturally leads
to an anti-correlation between the luminosity and duration,
providing an alternative interpretation for the Dainotti relation
\citep{Lenart:2025JHEAp..4700384L}.

Afterglows of short GRBs are generally poorly monitored,
particularly in terms of high-time-resolution and multi-wavelength
follow-up. Most short GRBs were only observed sporadically in the
optical band during the early afterglow phase, while radio
observations are even rarer \citep{Fong:2015ApJ...815..102F}. It
seriously limits both the precision of magnetar parameter
inference and the opportunity to systematically compare different
plateau origins. In addition, the small sample size introduces
significant statistical fluctuations in the
$\Gamma_0-E_{\gamma,\rm{iso}}$ correlation. Future high-cadence,
broadband follow-up observations will therefore be essential to
advance our understanding of short GRBs. The recently launched
Einstein Probe (EP) \citep{Yuan:2025SCPMA..6839501Y} and the
Space-based multiband astronomical Variable Objects Monitor (SVOM)
\citep{Wei:2016arXiv161006892W} missions are expected to
significantly enhance the multi-wavelength response capability and
substantially increase the sample size of GRB afterglows.
They will enable further test on the
$\Gamma_0-E_{\gamma,\rm{iso}}$ correlation in short GRBs and,
together with the EH--PS plane constructed from the
prompt-afterglow correlations \citep{Lenart:2025JHEAp..4700384L},
help clarify the origin of the plateau phase.

\section*{acknowledgements}
We thank the anonymous referee for helpful comments and
suggestions. This study was supported by the National Natural
Science Foundation of China (Grant Nos. 12233002, 12573051 and
12273113), by the National Key R\&D Program of China
(2021YFA0718500), and by the Major Science and Technology Program
of Xinjiang Uygur Autonomous Region (No. 2022A03013-1). YFH also
acknowledges the support from the Xinjiang Tianchi Program. JJG
acknowledges support from the Youth Innovation Promotion
Association (2023331). This work made use of data supplied by the
UK Swift Science Data Center at the University of Leicester.

\bibliography{sample631}{}
\bibliographystyle{aasjournal}

\renewcommand{\arraystretch}{1.5}
\begin{deluxetable*}{l c c c c c c c c c c}
\tabletypesize{\footnotesize} \label{table1}
\tablecaption{Model
parameters of the seven short GRBs derived from MCMC fitting of
the multi-band afterglow.}
\tablewidth{0pt}
\tablehead{
\colhead{GRB Name} & \colhead{$z$} & \colhead{$\log E_{\rm
K,iso}$} & \colhead{$\log \Gamma$} & \colhead{$\log n$} &
\colhead{$\log \theta_j$} & \colhead{$p$} &
\colhead{$\log \xi_e$} & \colhead{$\log \xi_B^2$} & \colhead{$\log B_p$} & \colhead{$\log P_0$} \\
\colhead{} & \colhead{} & \colhead{(erg)} & \colhead{} &
\colhead{$(\mathrm{cm^{-3}})$} & \colhead{(rad)} & \colhead{} &
\colhead{} & \colhead{} & \colhead{(G)} & \colhead{(s)}
}
\startdata
GRB 051221A       & 0.5464 & 52.04$^{+0.12}_{-0.11}$ & 3.25$^{+0.38}_{-0.46}$ & -2.04$^{+0.91}_{-0.60}$ & -1.37$^{+0.12}_{-0.07}$ & 2.04$^{+0.01}_{-0.01}$ & -0.57$^{+0.05}_{-0.10}$ & -2.46$^{+0.42}_{-0.58}$ & 14.43$^{+0.09}_{-0.08}$ & -2.96$^{+0.05}_{-0.03}$ \\
GRB 060614        & 0.125  & 49.52$^{+0.07}_{-0.02}$ & 2.90$^{+0.03}_{-0.07}$ &  -1.18$^{+0.14}_{-0.04}$ & -1.06$^{+0.02}_{-0.01}$ & 2.22$^{+0.01}_{-0.01}$ & -0.89$^{+0.10}_{-0.03}$ & -2.62$^{+0.11}_{-0.29}$ & 14.91$^{+0.03}_{-0.04}$ & -2.72$^{+0.02}_{-0.03}$ \\
GRB 070714B       & 0.923  & 51.43$^{+0.17}_{-0.15}$ & 2.98$^{+0.37}_{-0.44}$ & -1.59$^{+0.59}_{-0.62}$ & -1.66$^{+0.25}_{-0.21}$ & 2.47$^{+0.11}_{-0.11}$ & -0.57$^{+0.12}_{-0.15}$ & -1.54$^{+0.40}_{-0.46}$ & 14.97$^{+0.20}_{-0.80}$ & -2.74$^{+0.38}_{-0.14}$ \\
GRB 090510        & 0.903  & 52.16$^{+0.10}_{-0.08}$ & 3.38$^{+0.09}_{-0.13}$ & -2.41$^{+0.37}_{-0.46}$ & -2.02$^{+0.06}_{-0.07}$ & 2.18$^{+0.02}_{-0.02}$ & -0.58$^{+0.06}_{-0.11}$ & -1.51$^{+0.31}_{-0.25}$ & 12.60$^{+0.79}_{-0.45}$ & -2.34$^{+0.24}_{-0.36}$ \\
GRB 130603B       & 0.3568 & 52.11$^{+0.24}_{-0.14}$ & 2.23$^{+0.05}_{-0.04}$ & -0.90$^{+0.21}_{-0.24}$ & -1.52$^{+0.05}_{-0.05}$ & 2.06$^{+0.02}_{-0.02}$ & -0.85$^{+0.16}_{-0.19}$ & -2.62$^{+0.18}_{-0.20}$ & 15.10$^{+0.20}_{-0.52}$ & -2.53$^{+0.20}_{-0.19}$ \\
GRB 140903A       & 0.3529 & 51.51$^{+0.15}_{-0.14}$ & 2.63$^{+0.38}_{-0.29}$ & -2.62$^{+0.36}_{-0.26}$ & -1.60$^{+0.05}_{-0.04}$ & 2.03$^{+0.01}_{-0.01}$ & -0.51$^{+0.08}_{-0.13}$ & -1.78$^{+0.22}_{-0.29}$ & 14.82$^{+0.10}_{-0.07}$ & -2.97$^{+0.03}_{-0.02}$ \\
GRB 150424A      & 0.3 & 51.05$^{+0.08}_{-0.08}$ & 2.18$^{+0.03}_{-0.03}$ & -3.06$^{+0.06}_{-0.03}$ & -1.43$^{+0.02}_{-0.02}$ & 2.05$^{+0.002}_{-0.002}$ & -0.51$^{+0.00}_{-0.01}$ & -1.55$^{+0.04}_{-0.08}$ & 14.75$^{+0.05}_{-0.05}$ & -2.76$^{+0.03}_{-0.03}$ \\
\enddata
\end{deluxetable*}

\begin{figure}[ht!]
\centering
\gridline{
  \fig{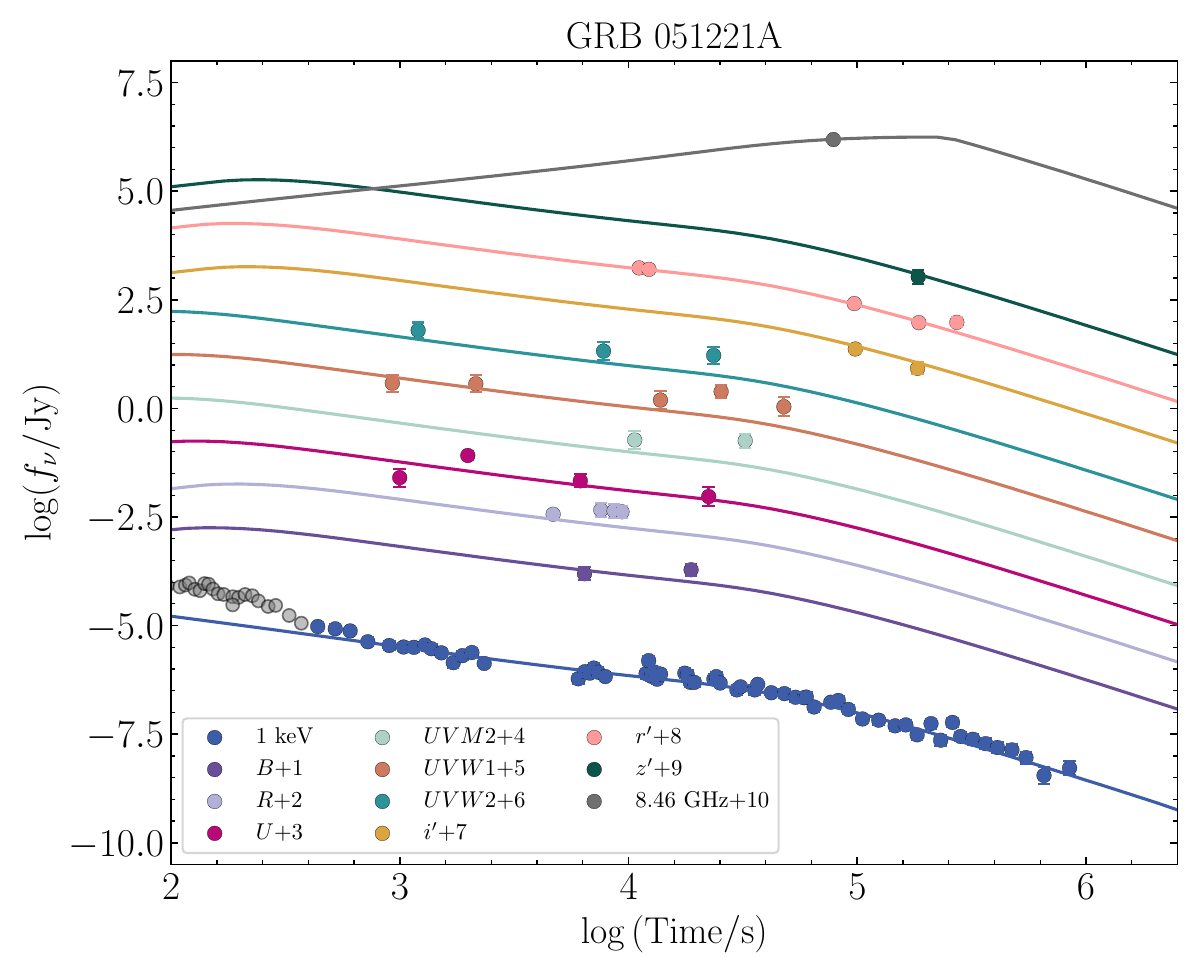}{0.45\textwidth}{} \hspace{-0.5cm}
  \fig{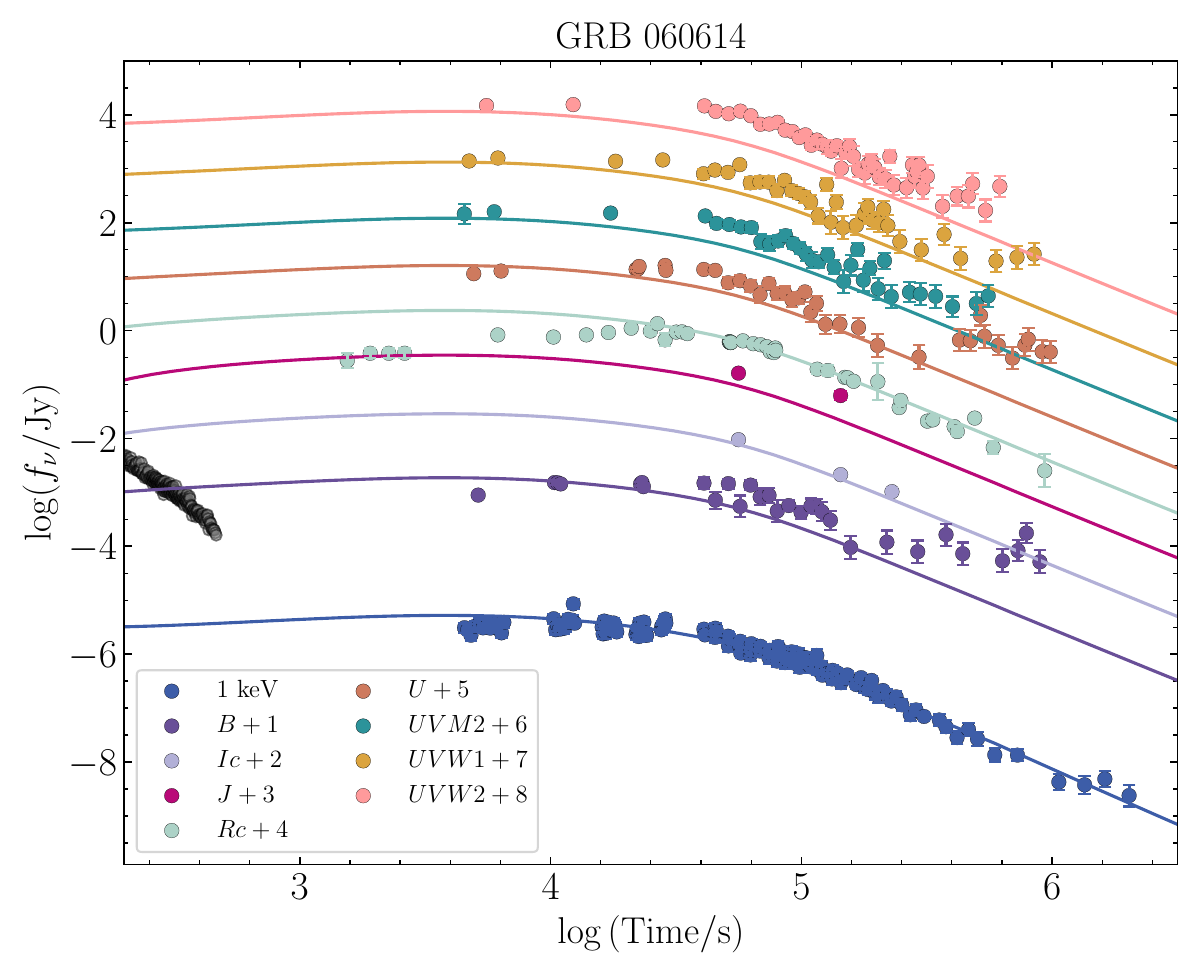}{0.45\textwidth}{}
}
\vspace{0cm}
\gridline{
  \fig{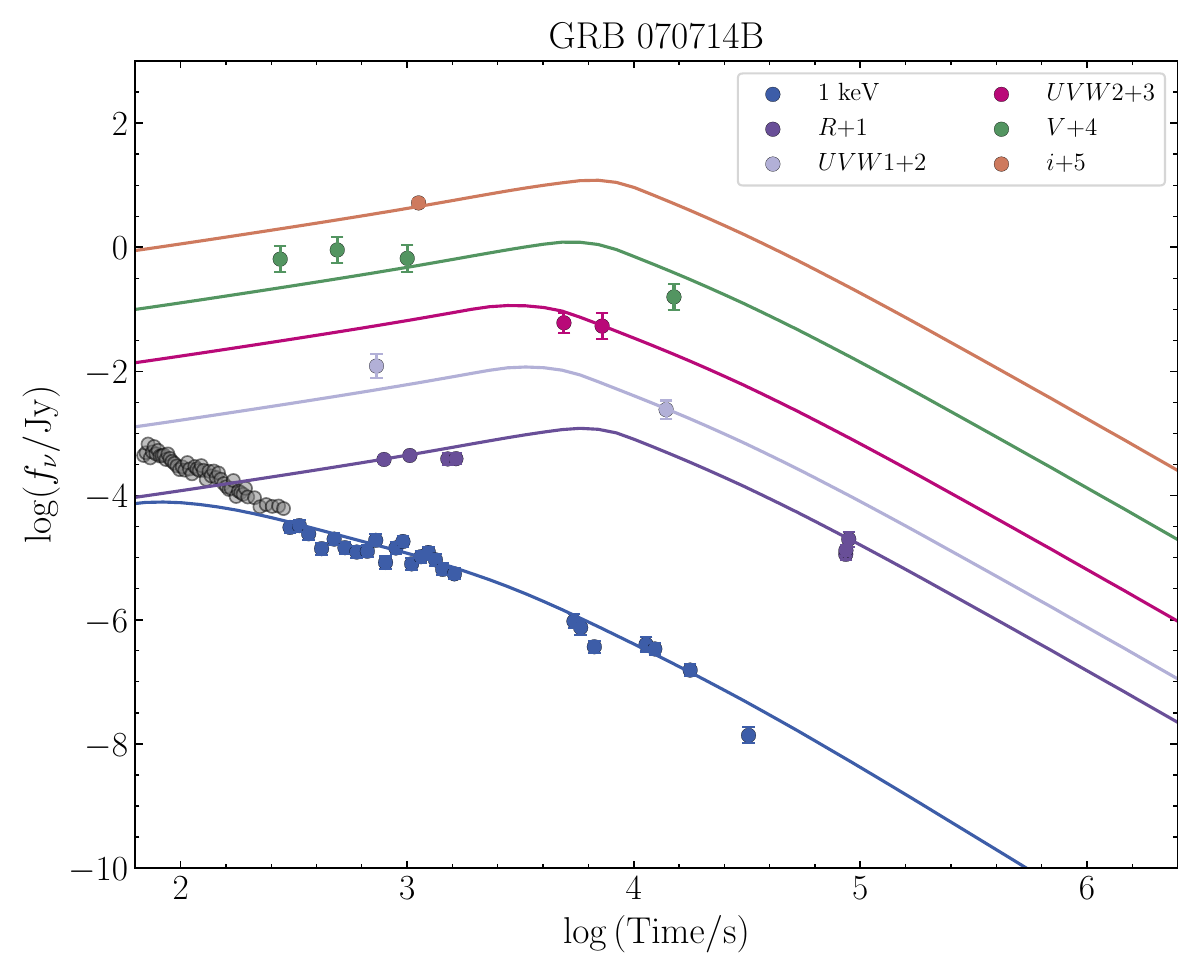}{0.45\textwidth}{} \hspace{-0.5cm}
  \fig{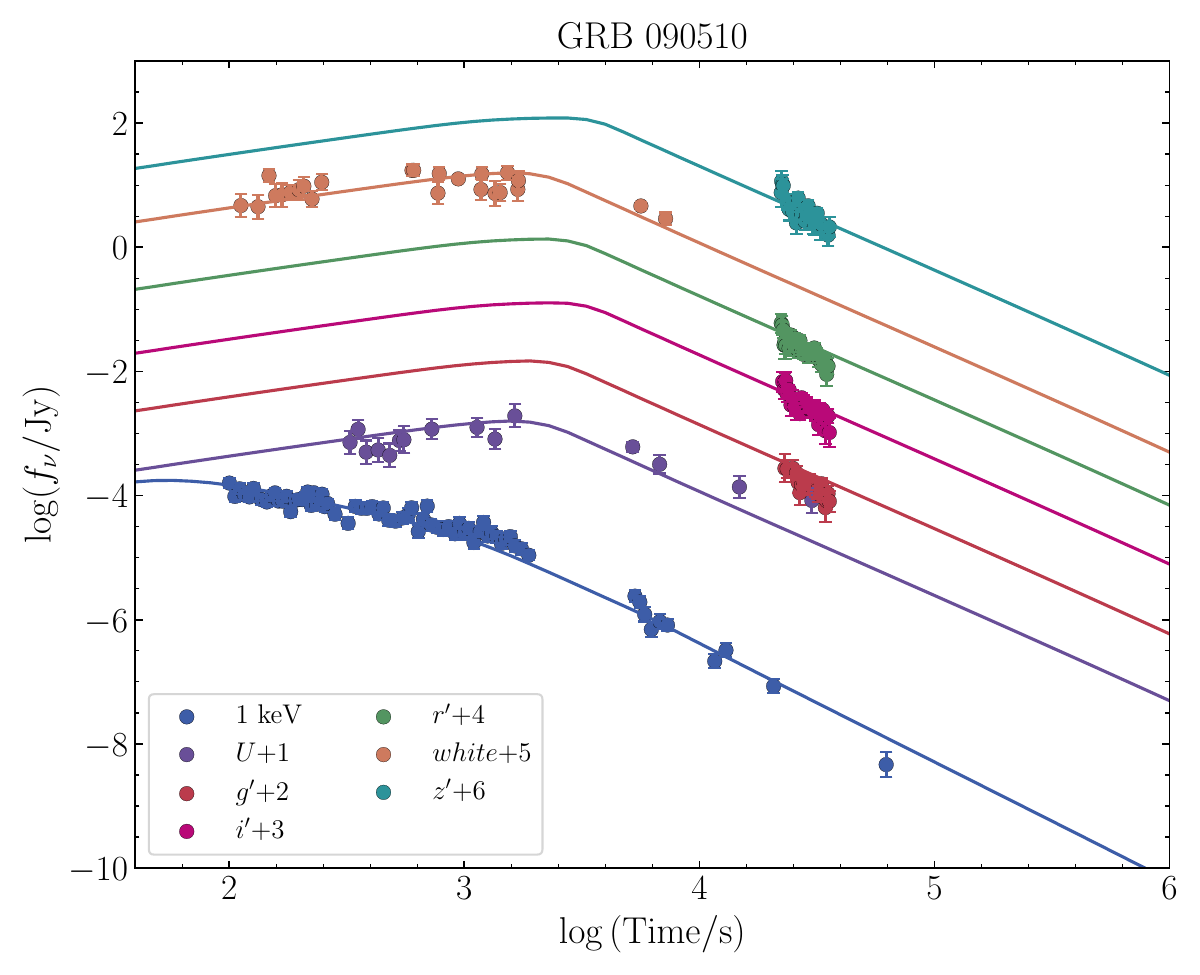}{0.45\textwidth}{}
} \vspace{0cm}
\caption{Multi-wavelength afterglow light curves of seven short
GRBs and the best fit results in the framework of the magnetar
energy injection scenario. For our sample, X-ray data are taken
from the \textit{Swift} light curve repository
\citep{Evans:2007A&A...469..379E,Evans:2009MNRAS.397.1177E},
optical data from the optical GRB catalogue compiled by
\cite{Dainotti:2024MNRAS.533.4023D}, and radio data from the
published literature and the General Coordinates Network
(see Section \ref{sec4} for details). The gray semi-transparent
data points may be affected by late prompt emission or extended
emission and are thus excluded from the fitting. For clarity,
light curves at different frequencies have been vertically
shifted.} \label{fig2}
\end{figure}

\begin{figure}[ht!]
\centering
\figurenum{2}
\gridline{
  \fig{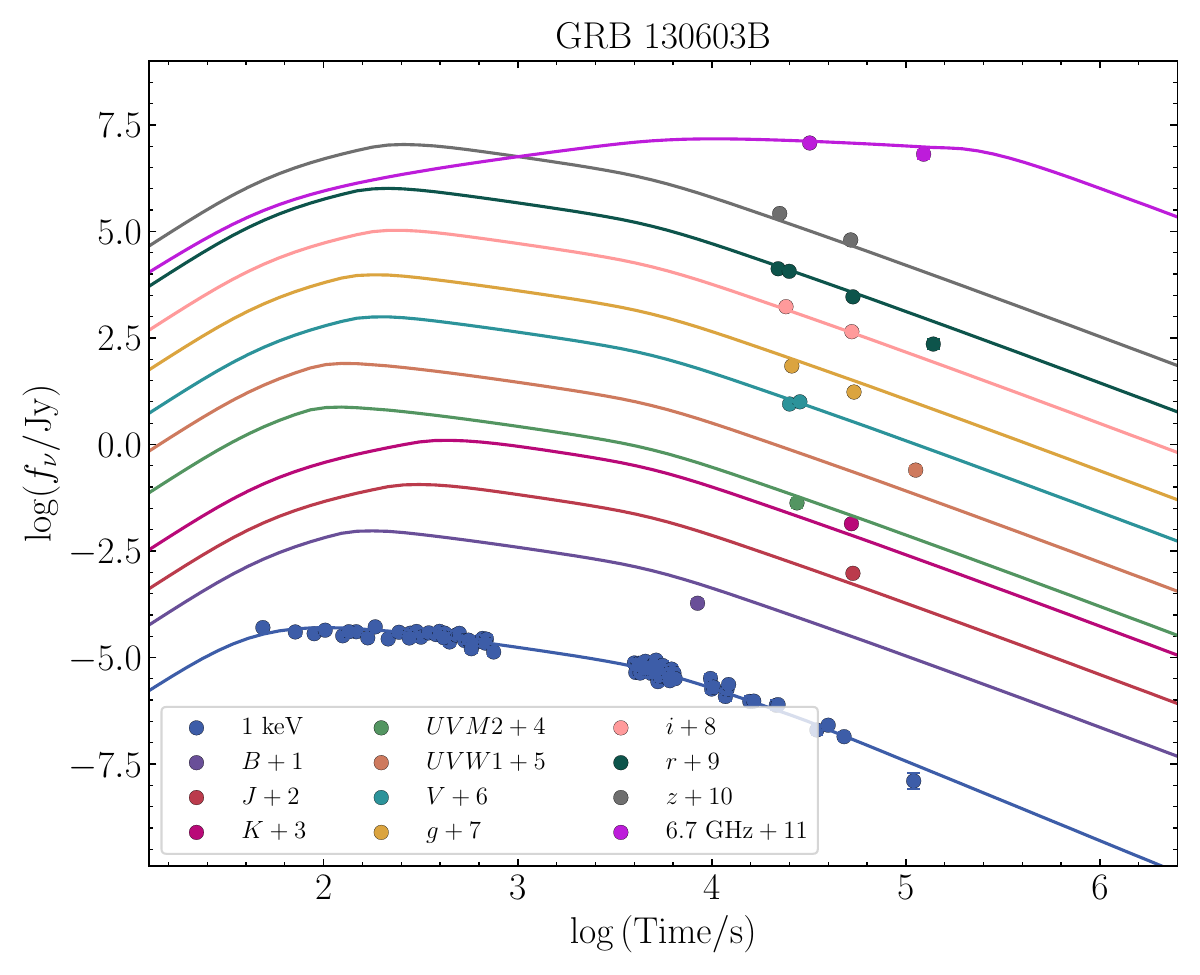}{0.45\textwidth}{} \hspace{-0.5cm}
  \fig{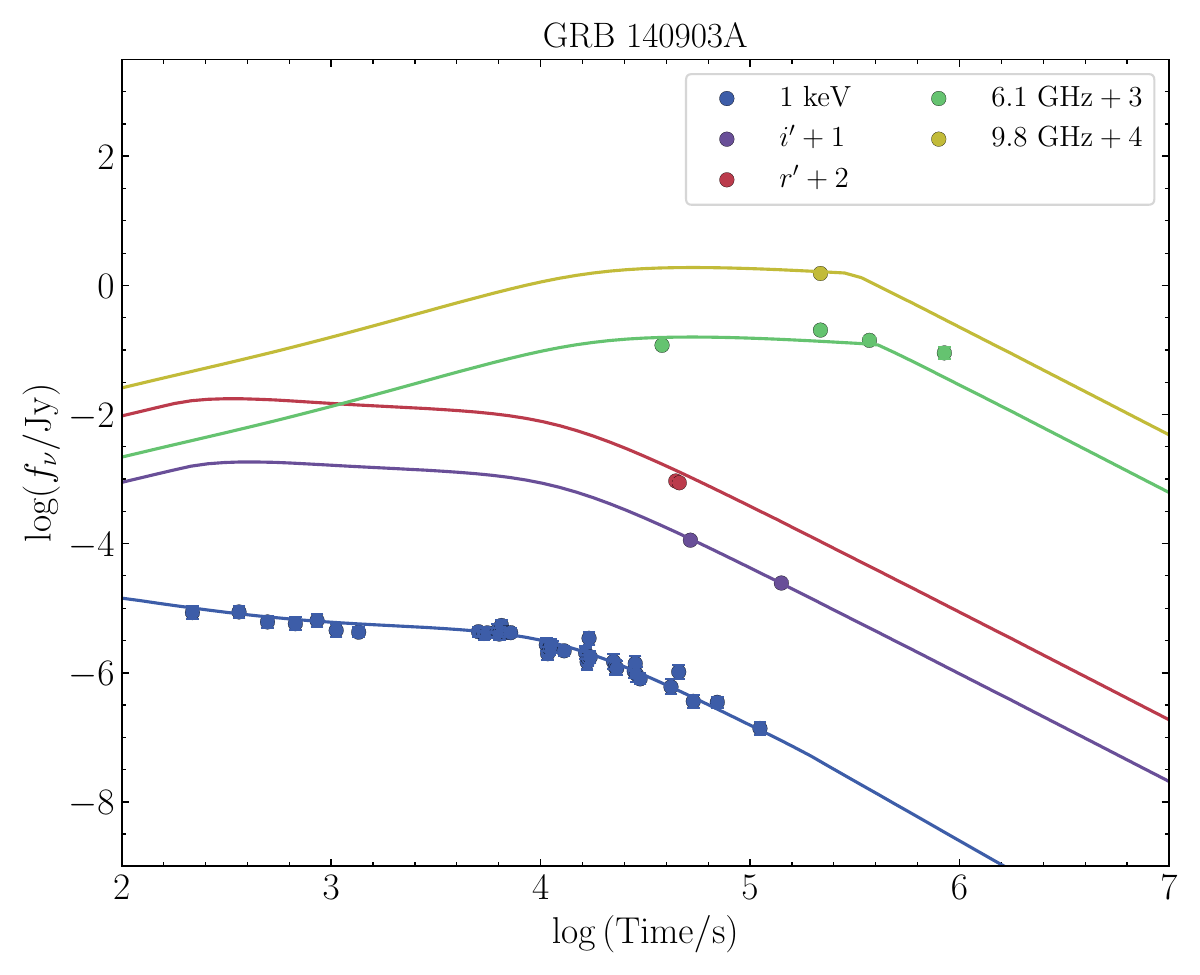}{0.45\textwidth}{}
}
\vspace{0cm}
\gridline{
  \fig{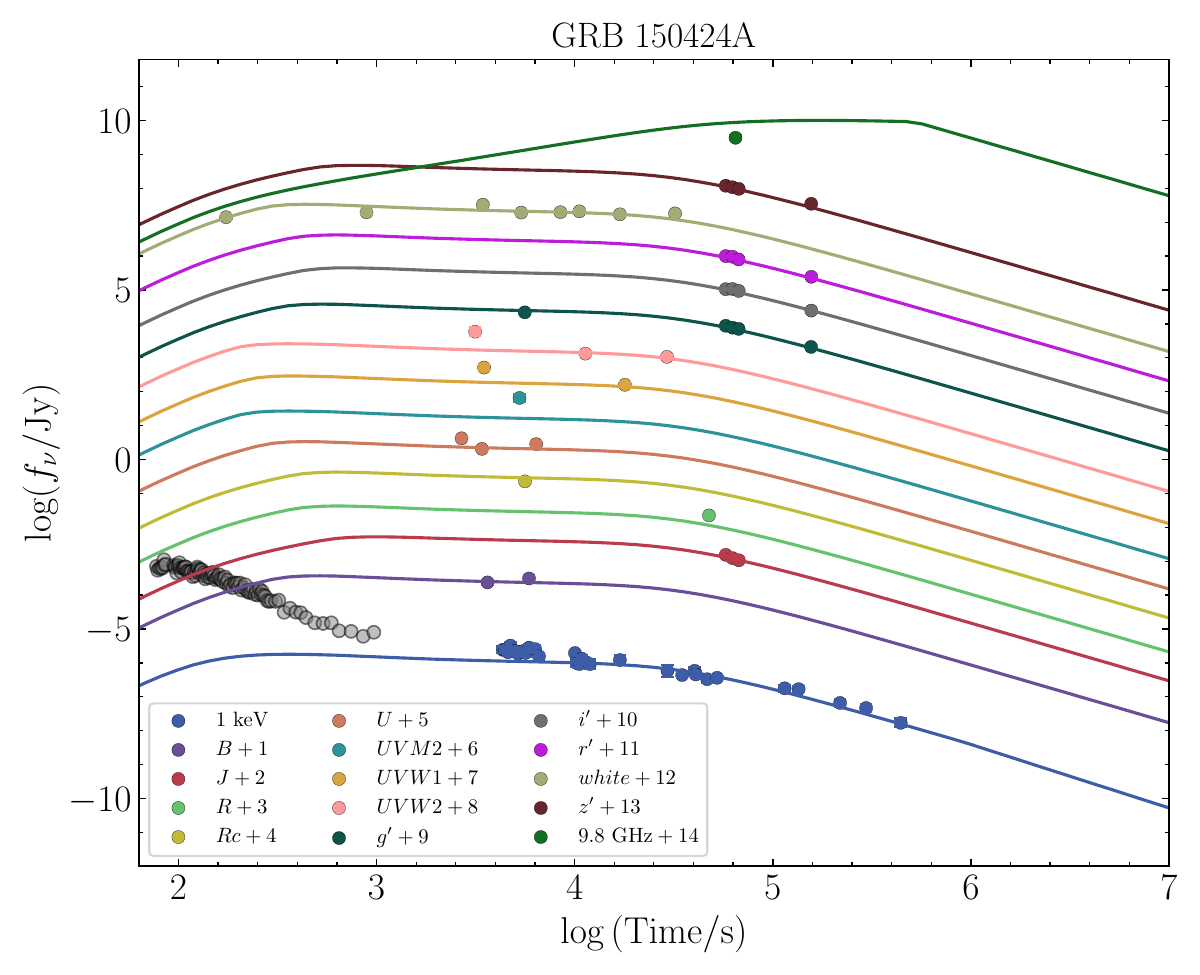}{0.45\textwidth}{} \hspace{-0.5cm}
}
\caption{(Continued.)}
\end{figure}

\begin{figure}[ht!]
\centering
\gridline{
  \fig{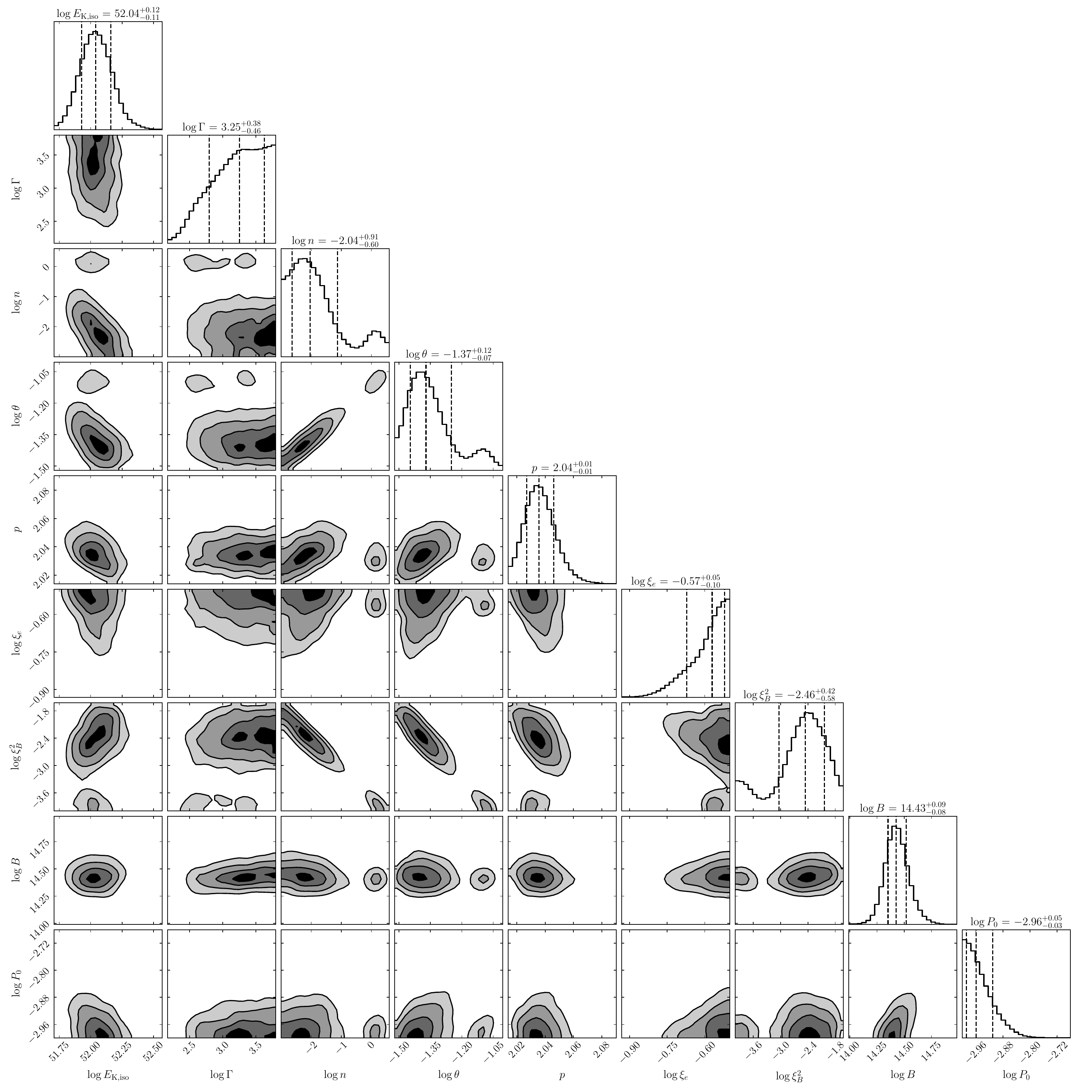}{0.45\textwidth}{}\hspace{-0.5cm}
  \fig{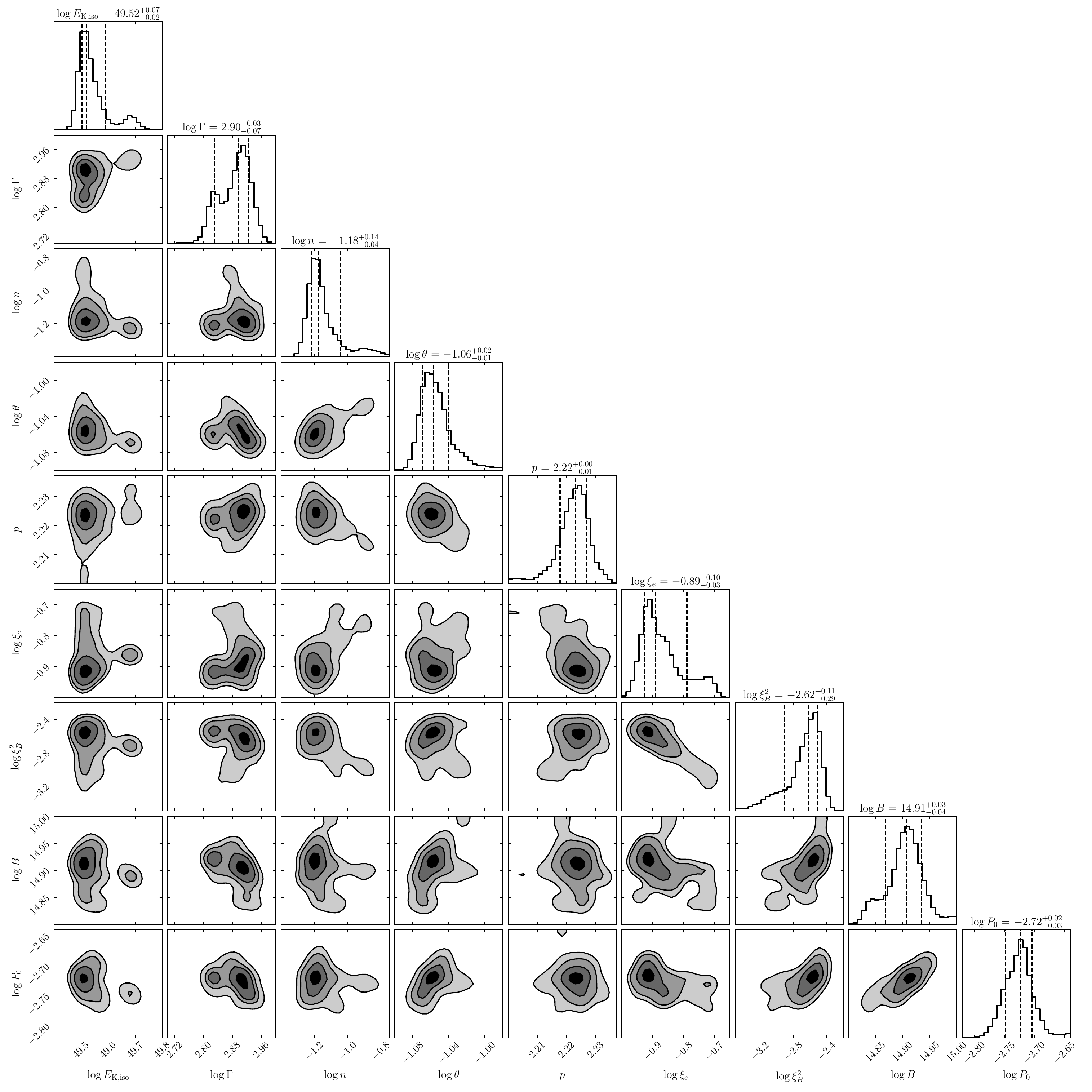}{0.45\textwidth}{}
}
\centerline{(a)\hspace{8cm}(b)}
\vspace{0cm}
\gridline{
  \fig{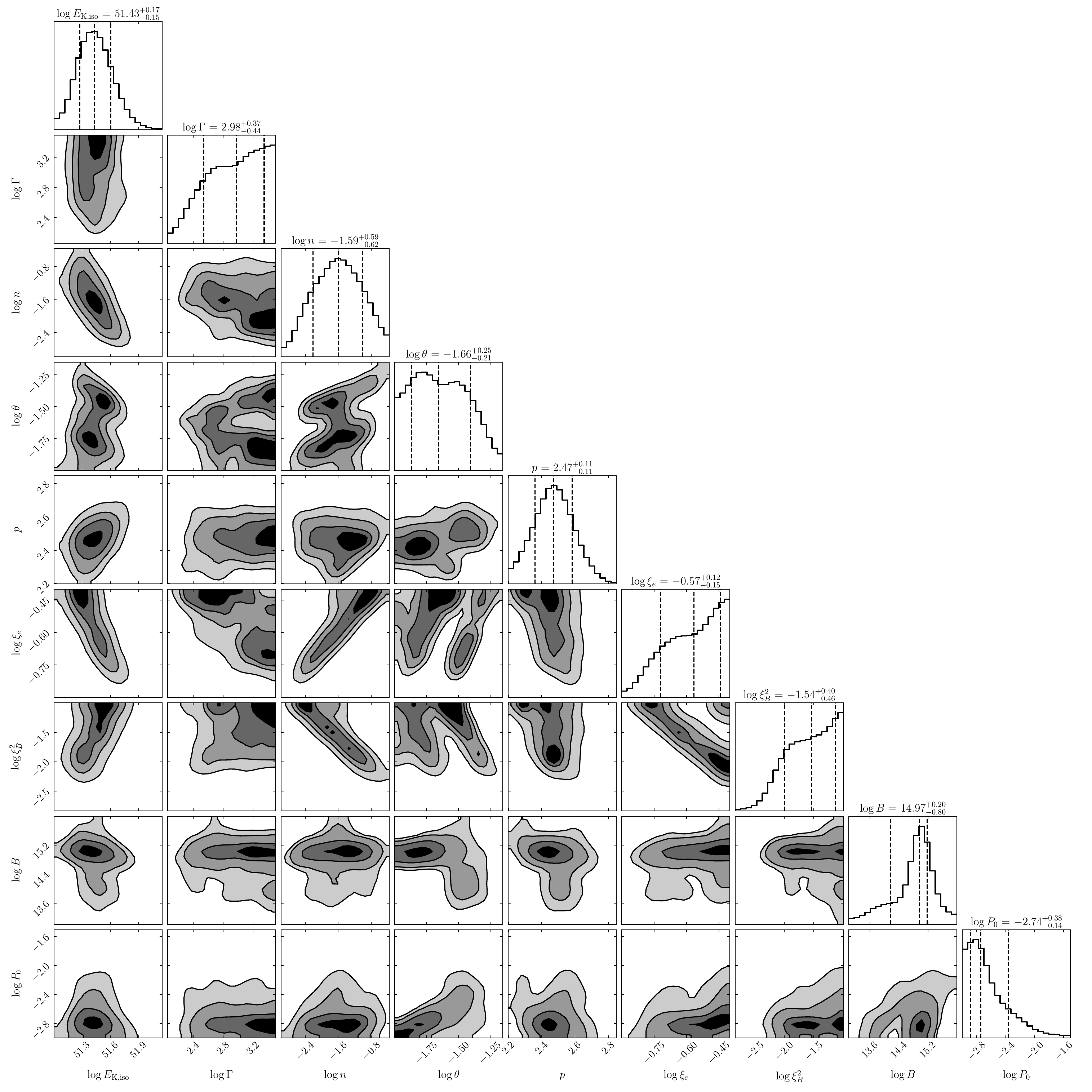}{0.45\textwidth}{} \hspace{-0.5cm}
  \fig{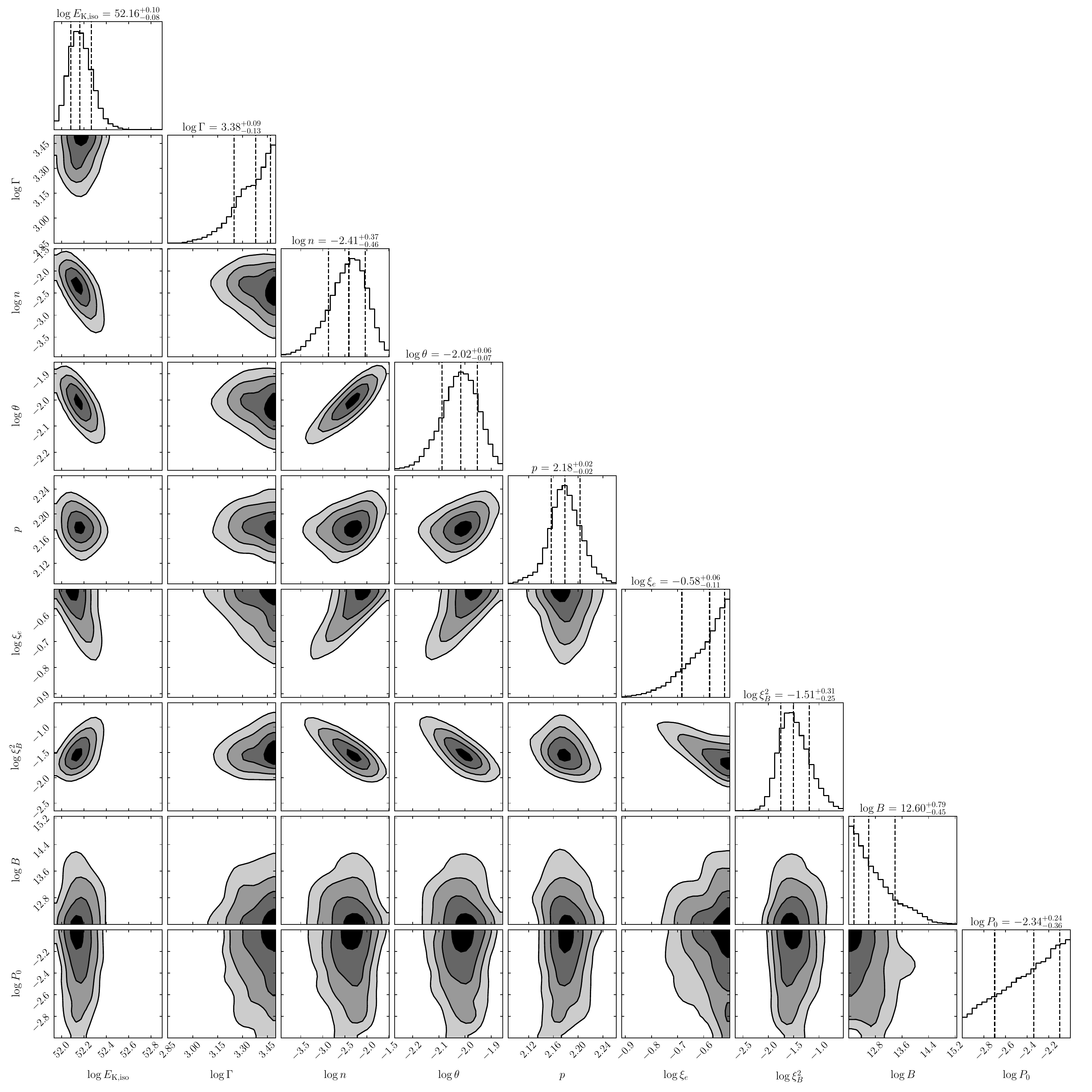}{0.45\textwidth}{}
} \centerline{(c)\hspace{8cm}(d)} \vspace{0cm}
 \caption{Posterior
probability distributions of the model parameters derived from
MCMC fitting for the seven short GRBs. The dashed lines indicate
the 16th, 50th, and 84th percentiles, corresponding to the median
and $1\sigma$ confidence intervals. Panels (a)--(h) correspond to
GRBs 051221A, 060614, 070714B, 090510, 130603B, 140903A, and 150424A,
respectively.} \label{fig3}
\end{figure}

\begin{figure}[ht!]
\centering
\figurenum{3}
\gridline{
  \fig{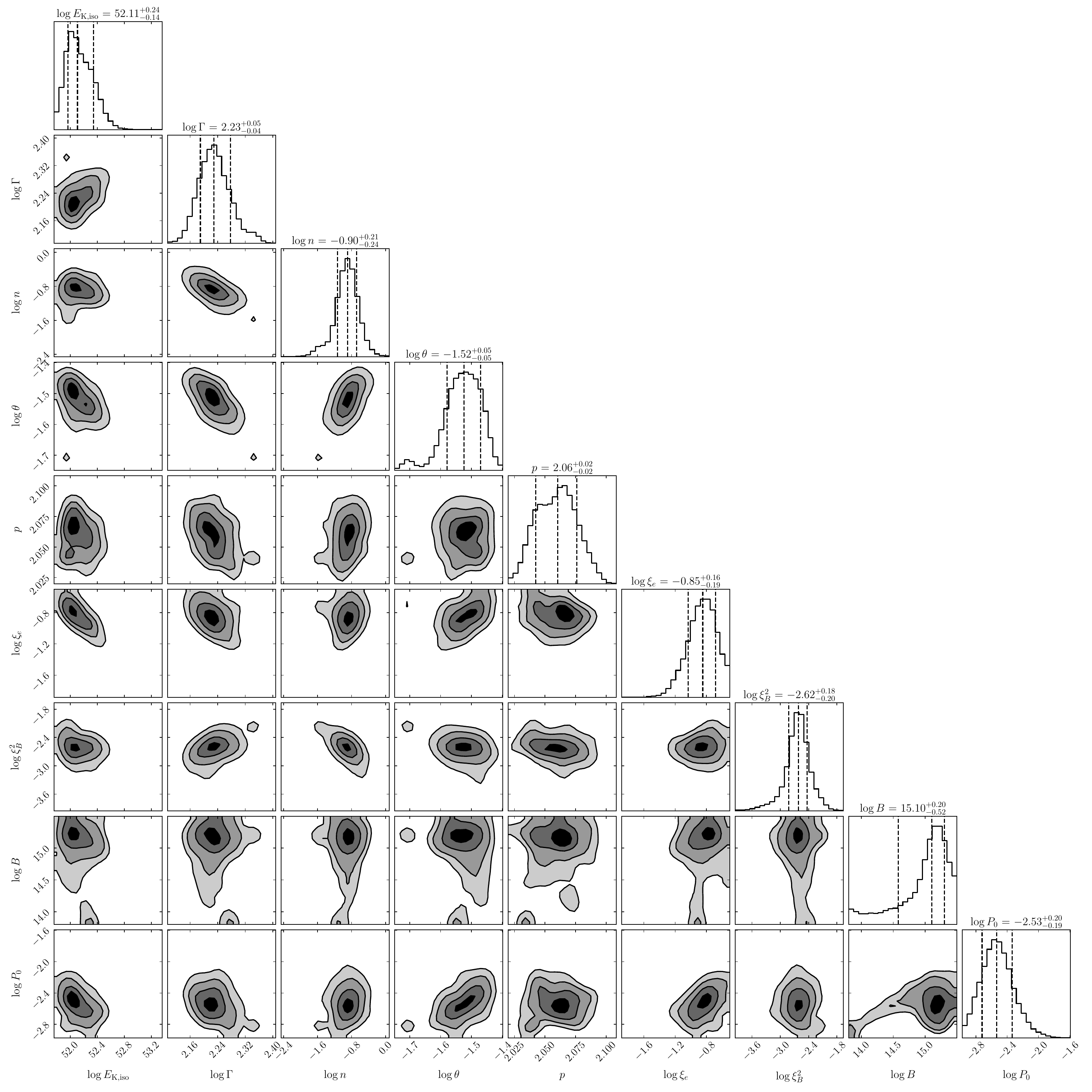}{0.45\textwidth}{} \hspace{-0.5cm}
  \fig{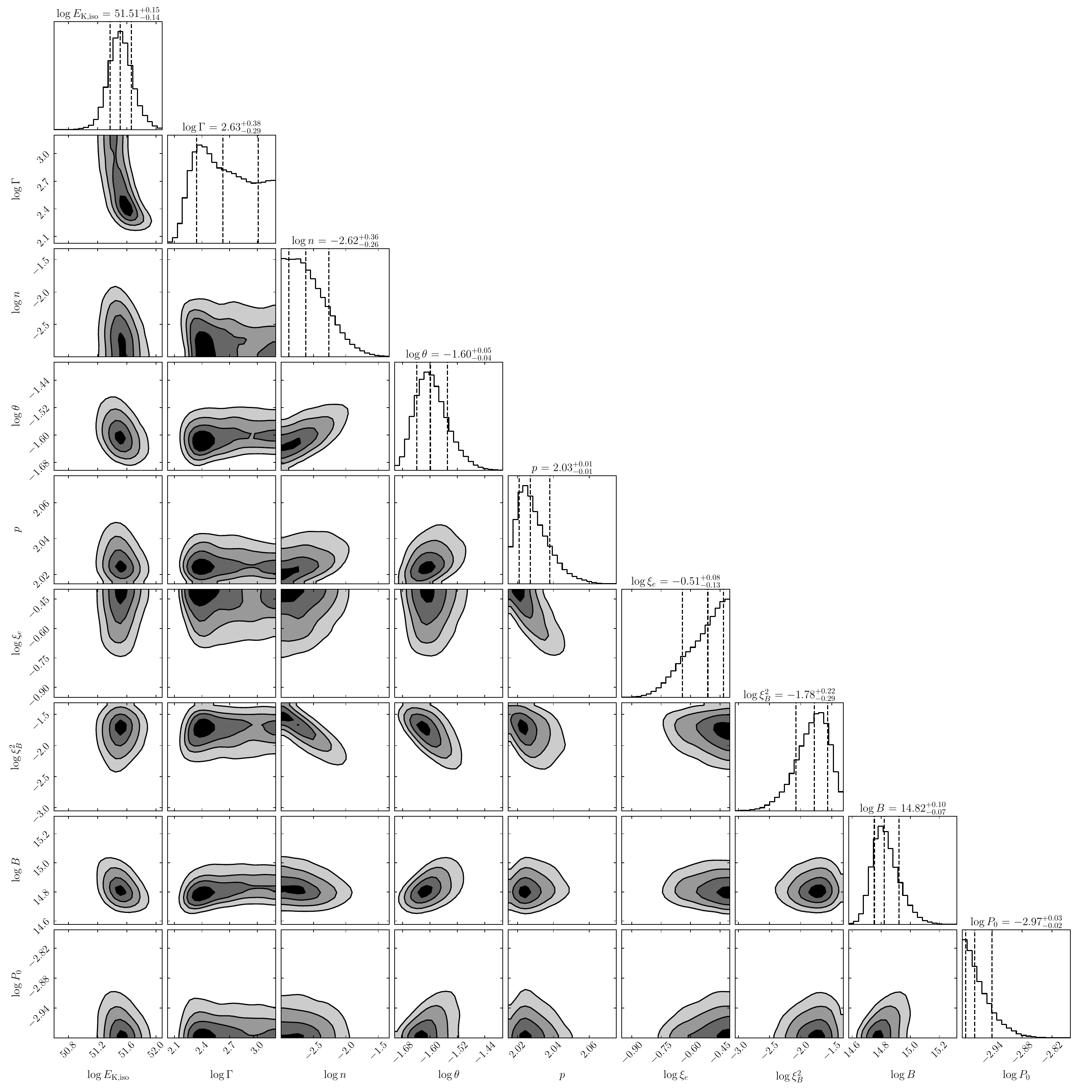}{0.45\textwidth}{}
}
\centerline{(e)\hspace{8cm}(f)}
\vspace{0cm}
\gridline{
  \fig{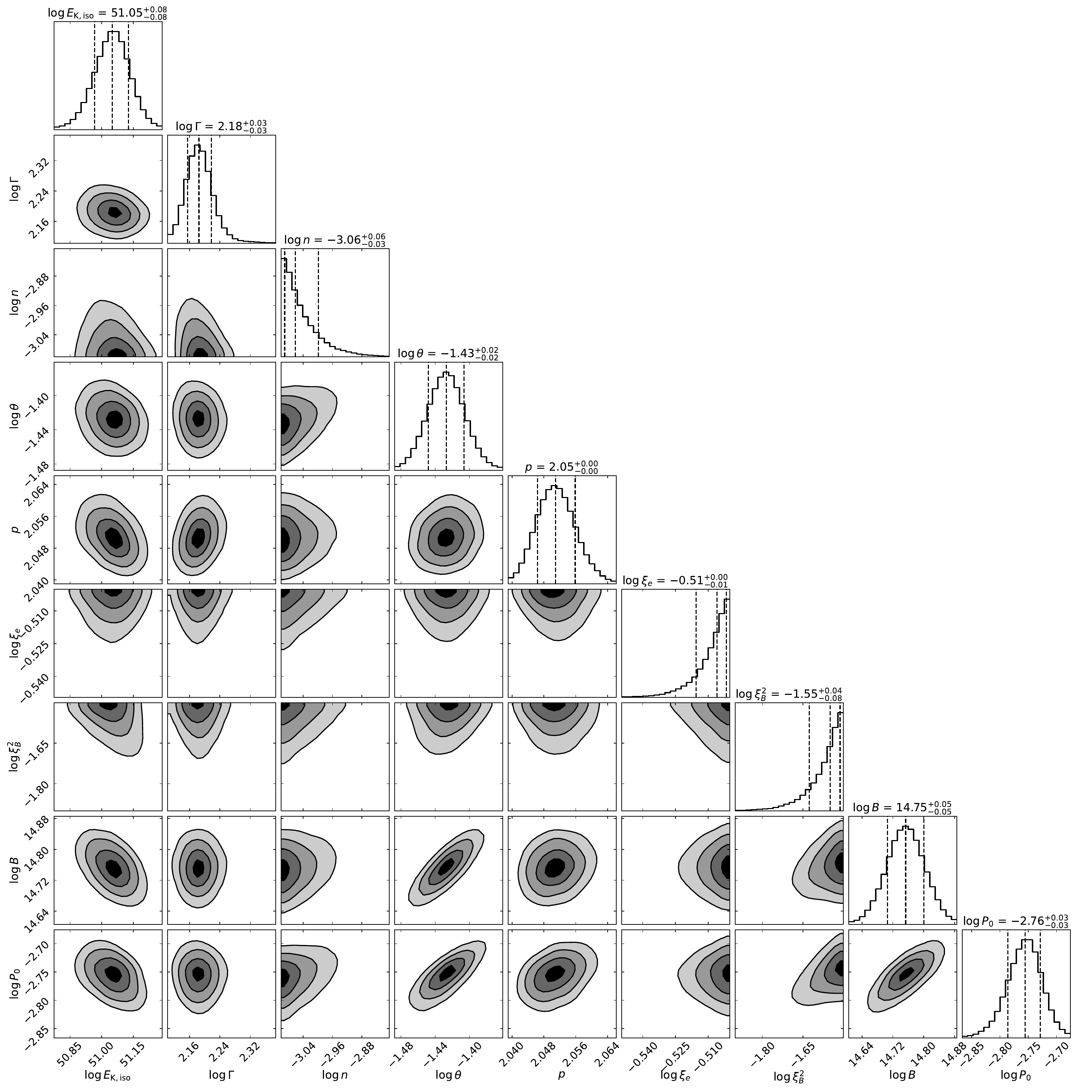}{0.45\textwidth}{} \hspace{-0.5cm}
}
\centerline{(g)}
\caption{(Continued.)}
\end{figure}

\begin{figure}[ht!]
\gridline{\fig{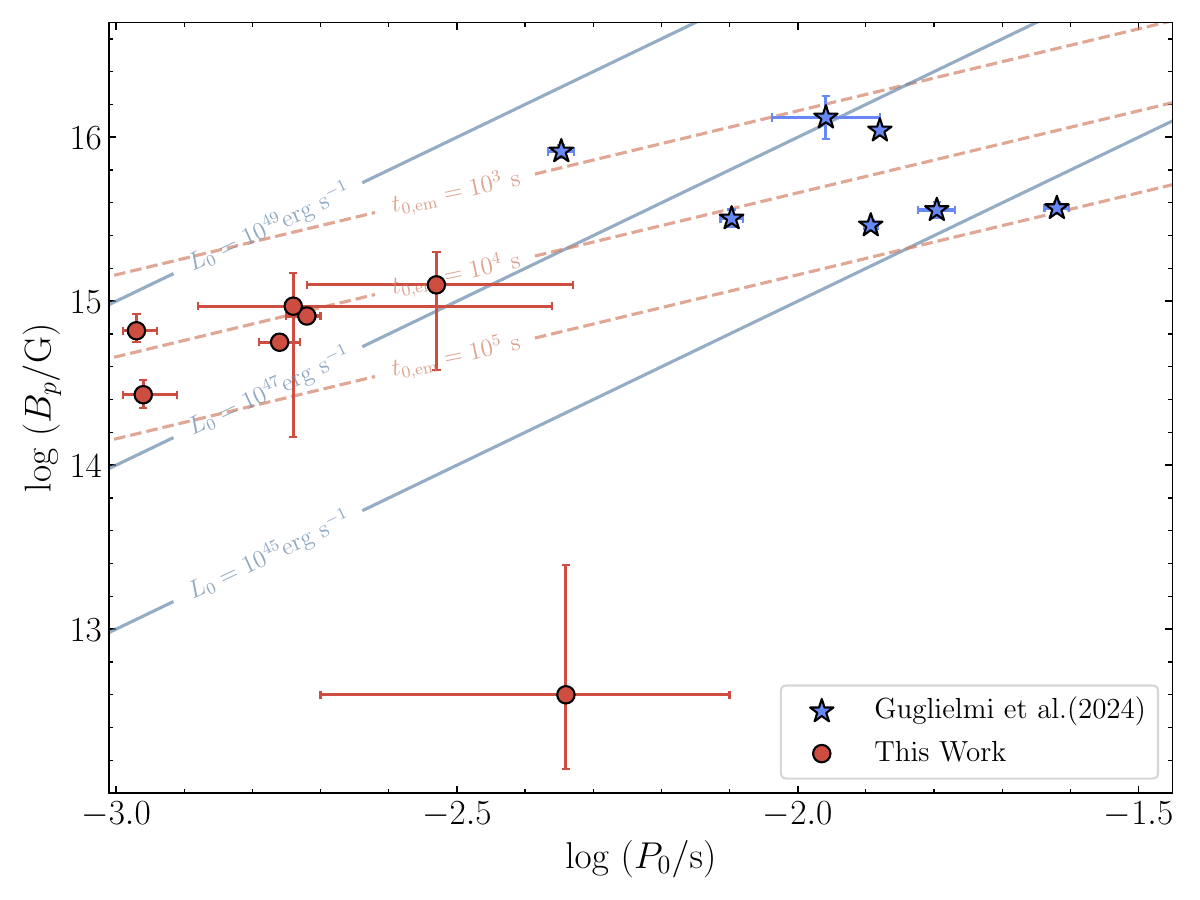}{0.6\textwidth}{}}
\caption{The dipole magnetic field strength $B_p$ versus the
initial spin period $P_0$ of the central magnetar for the seven
short GRBs. The circles represent our best-fit results obtained
from multi-wavelength afterglow modeling, with error bars
indicating $1\sigma$ uncertainties. The star symbols indicate
the corresponding parameters derived by \cite{Guglielmi:2024A&A...692A..73G}
based on X-ray afterglow modeling.
The short GRB sample of \citet{Stratta:2018ApJ...869..155S}
is also included for comparison, shown as diamond symbols.
The solid and dashed lines represent the loci of
constant initial dipole luminosity $L_0$ and characteristic
spin-down timescale $t_{0,\mathrm{em}}$, respectively.
\label{fig4}}
\end{figure}

\begin{figure}[ht!]
\gridline{\fig{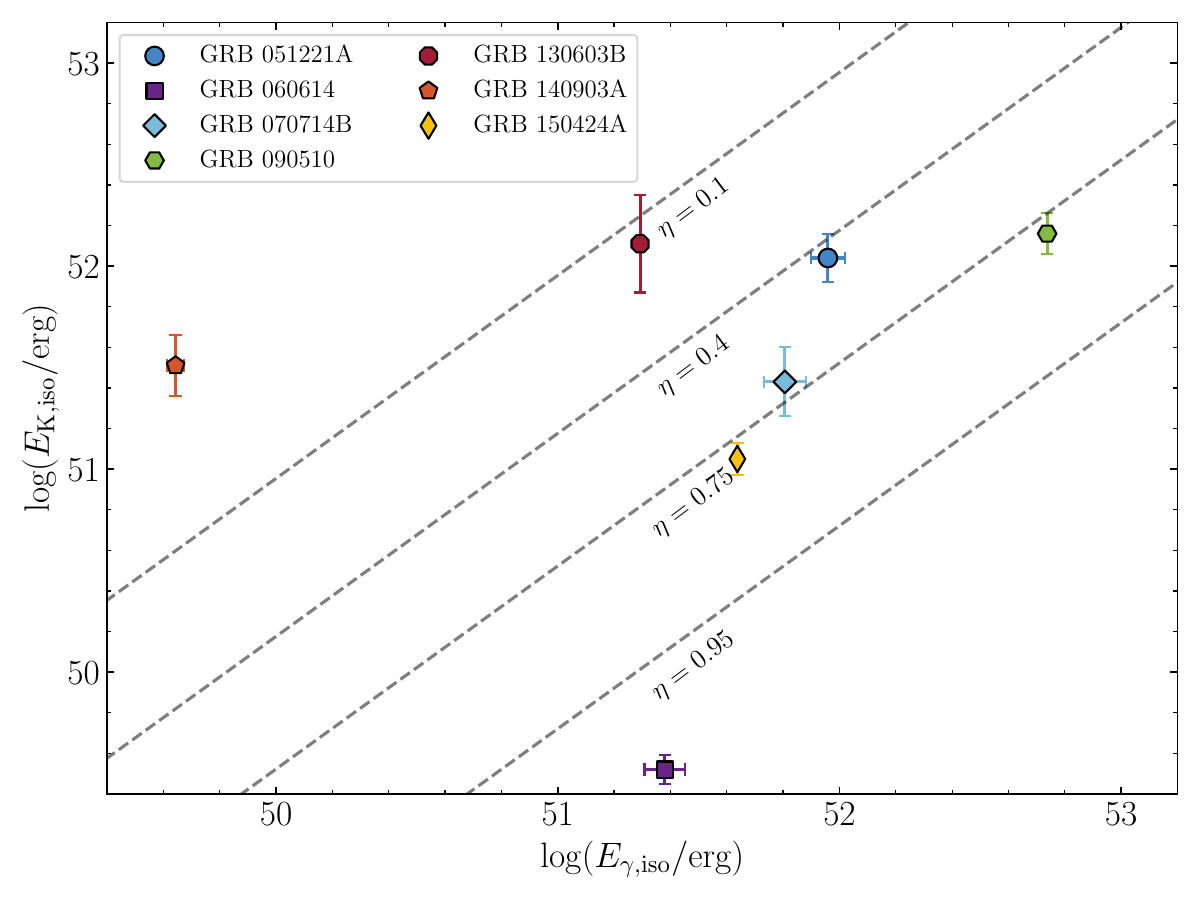}{0.6\textwidth}{}}
 \caption{The
isotropic-equivalent kinetic energy $E_{\mathrm{K,iso}}$ plotted
versus the gamma-ray energy $E_{\gamma,\mathrm{iso}}$ for the
seven short GRBs. The dashed lines represent constant radiative
efficiencies $\eta_{\gamma} = E_{\gamma,\mathrm{iso}} /
(E_{\gamma,\mathrm{iso}} + E_{\mathrm{K,iso}})$ with values of
0.1, 0.4, 0.75, and 0.95. \label{fig5}}
\end{figure}

\begin{figure}[ht!]
\gridline{\fig{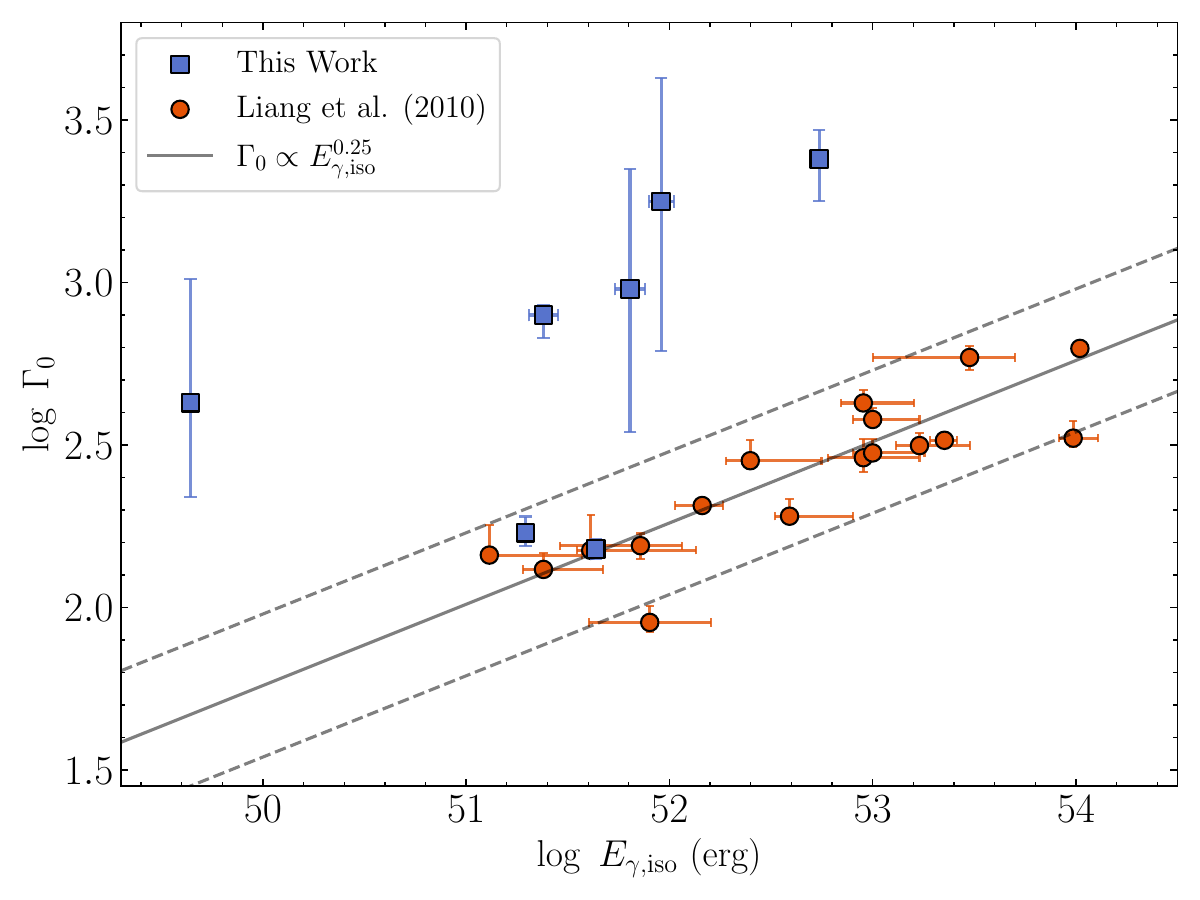}{0.6\textwidth}{}}
\caption{Correlation between the initial Lorentz factor $\Gamma_0$
and the isotropic-equivalent gamma-ray energy $E_{\gamma,\mathrm{iso}}$.
The blue squares represent the best-fit results for the seven short
GRBs obtained from multi-wavelength afterglow modeling in this study,
while red circles denote the long GRB sample of
\cite{Liang:2010ApJ...725.2209L}. The solid line shows the
best-fit $\Gamma_0-E_{\gamma,\mathrm{iso}}$ correlation, and the
dashed lines indicate the $2\sigma$ confidence region, with
$\sigma=0.11$.
\label{fig6}}
\end{figure}

\end{document}